\documentclass[11pt]{article}
\usepackage{amssymb,amsthm,amsmath,float,subfigure}
\usepackage{appendix}
\usepackage{booktabs} 

\usepackage[total={6.5in,8.75in}, top=1.2in, left=0.9in, includefoot]{geometry}
\usepackage{graphicx}
\newcommand{\Eq}[1]{(\ref{eq:#1})}

\newcommand{\Lem}[1]{Lem.~\ref{lem:#1}}

\newcommand{\Sec}[1]{\S \ref{sec:#1}}
\newcommand{\Fig}[1]{Fig.~\ref{fig:#1}}
\newcommand{\Tbl}[1]{Table~\ref{tbl:#1}}

\newcommand{\InsertFig}[4]
{\begin{figure}[h!t]
 \centerline{
 \includegraphics[width=#4]{#1}
 }
 \caption{{\footnotesize #2}
 \label{fig:#3}}
\end{figure}}

\newcommand{\InsertFigTwo}[5] {
\begin{figure}[h!t]
 \centerline{
 \includegraphics[width=#5]{#1}
 \hskip 0.1in
 \includegraphics[width=#5]{#2}
 }
 \caption{{\footnotesize #3}
 \label{fig:#4}}
\end{figure}}

\newcommand{\bC}{{\mathbb{ C}}}
\newcommand{\bN}{{\mathbb{ N}}}

\newcommand{\bQ}{{\mathbb{ Q}}}
\newcommand{\bR}{{\mathbb{ R}}}
\newcommand{\bS}{{\mathbb{ S}}}
\newcommand{\bT}{{\mathbb{ T}}}
\newcommand{\bZ}{{\mathbb{ Z}}}

\newcommand{\cD}{{\cal D}}

\newcommand{\cO}{{\cal O}}

\newcommand{\cS}{{\cal S}}
\newcommand{\cT}{{\cal T}}

\newcommand{\eps}{\varepsilon}
\newcommand{\vphi}{\varphi}

\newtheorem{thm}{Theorem}
\newtheorem{lem}[thm]{Lemma}

\newcommand{\beq}[1]{\begin{equation}\label{eq:#1}}
\newcommand{\eeq}{\end{equation}}

\newenvironment{se}[1]{\equation\label{eq:#1}\aligned}{\endaligned\endequation}
\newcommand{\bsplit}[1]{\begin{se}{#1}}
\newcommand{\esplit}{\end{se}}

\newenvironment{example}[1][]
 {
	\setlength \leftmargini {1.0em}		
	\setlength \topsep {0.5em}			
	\begin{quote}
	{\it Example#1} }
	{\end{quote}
 }
\newcommand{\bexam}[1][:]{\begin{example}[#1]}
\newcommand{\eexam}{\end{example}}

\title{Efficient Computation of Invariant Tori in Volume-Preserving Maps}
\author{Adam M.~Fox\footnote{Adam.Fox@colorado.edu}
 \ and\ 
 James D.~Meiss\footnote{James.Meiss@colorado.edu}
	\ \thanks
 {
 AMF and JDM and were supported in part by NSF grant DMS-1211350. 
 Useful conversations with Timothy Blass, Rafael de la Llave, Keith Julien, 
 and Holger Dullin are gratefully acknowledged. 
 }
 \\
 	Department of Applied Mathematics\\
 	University of Colorado \\
	Boulder, CO 80309-0526 \\
}
\date{\today}
\begin{document}
\maketitle

\begin{abstract}
\vspace*{1ex}
\noindent

In this paper we implement a numerical algorithm to compute codimension-one tori in three-dimensional, volume-preserving maps. A torus is defined by its conjugacy to rigid rotation, which is in turn given by its Fourier series. The algorithm employs a quasi-Newton scheme to find the Fourier coefficients of a truncation of the series. This technique is based upon the theory developed in the accompanying article \cite{Blass13}. It is guaranteed to converge assuming the torus exists, the initial estimate is suitably close, and the map satisfies certain nondegeneracy conditions. We demonstrate that the growth of the largest singular value of the derivative of the conjugacy predicts the threshold for the destruction of the torus.  We use these singular values to examine the mechanics of the breakup of the tori, making comparisons to Aubry-Mather and anti-integrability theory when possible.

\end{abstract}

\section{Introduction}\label{sec:Intro}
In this paper we will study families of maps $f_{\eps}$ on $M=\bT^2 \times \bR$ of the form
\bsplit{NearIntegrable}
	x' &= x + \Omega(z) - \eps g_1(x,z,\eps) \mod 1 ,\\
	z' &= z - \eps g_2(x,z,\eps).
\esplit
where $x\in \bT^2$ are period-one \emph{angles} and $z\in\bR$ represents the \emph{action}. Maps of this form arise in the study of many natural phenomena including incompressible fluids \cite{Piro88, Finn88b, Cartwright96}, magnetic field-lines \cite{Thyagaraja85}, granular mixing \cite{Meier07}, and celestial mechanics \cite{Liu94}. 

Invariant tori play a prominent role in the dynamics of maps such as \Eq{NearIntegrable}. Most notably, when $\eps=0$ every orbit lies on a two-torus $\cT_{z_0}=\bT^2 \times \{z_0\}$ on which the dynamics is simply a rigid rotation 
\beq{RigidRot}
	(x_t,z_t)=(x_0+\omega t, z_0), \quad \omega = \Omega(z_0);
\eeq
such maps are \emph{integrable}. 

We will assume that \Eq{NearIntegrable} is an exact volume-preserving map $f^*\alpha - \alpha = dS$, with respect to the two-form $\alpha = z dx_1 \wedge dx_2$. Exactness is a necessary condition for the existence of \emph{rotational} invariant tori when $\eps \neq 0$. We say a torus is \emph{rotational} if it is homotopic to $\cT_0$. The rotational tori of \Eq{NearIntegrable} are fundamentally important as they form boundaries to transport: orbits that begin on one side cannot cross to the other \cite[App. B]{Feingold88}. 

KAM theory \cite{Cheng90a, Xia92, Blass13} guarantees the persistence of a Cantor set of rotational tori in exact volume-preserving, near-integrable maps of the form \Eq{NearIntegrable}, assuming that $g_1$, $g_2$, and $\Omega$ are smooth enough, and the frequency map $\Omega: \bR \to \bR^2$ satisfies a Kolmogorov nondegeneracy assumption
\[
	\mbox{rank}(\Omega, D\Omega, D^2\Omega \ldots) \ge 2 .
\]
The persistent tori have rotation vectors that are \emph{Diophantine},
\beq{Diophantine}
	\cD_s = \left\{\omega \in \bR^2| \exists\, c > 0 \;\;\mbox{s.t. } |p\cdot \omega-q| > {c}{|p|^{-s}}\right\},
\eeq
for some $s \ge 2$. It is important to note that, unlike the Hamiltonian case, this theory does not predict the persistence of a torus with a given rotation vector for a given map $f_\eps$, but just a Cantor set of Diophantine tori when $\eps$ is small enough.

Although KAM theory does guarantee the persistence of some of tori for small $\eps$, it does not say anything about what happens when $\eps = \cO(1)$. John Greene \cite{Greene79} developed the first quantitative method to study the persistence of tori for the case of two-dimensional, area-preserving maps. He conjectured that periodic orbits in the neighborhood of an invariant circle should be stable; indeed, a sequence of periodic orbits should limit upon a circle only if they remained stable in the limit. Conversely, if the limit of a family of periodic orbits is unstable, then the invariant circle should no longer exist. This method is known as \emph{Greene's residue criterion} \cite{Greene79}. In \cite{Fox13a}, we extended this method to the volume-preserving case; however, our generalization is only applicable to maps with reversing symmetries and symmetric tori.

In this paper, we will show how the reversibility requirement can be circumvented by studying the tori directly, rather than a sequence of approximating periodic orbits. In \Sec{FourierMethod} we describe a Fourier-based scheme (developed in conjunction with Blass and de la Llave \cite{Blass13}) to compute the embedding for an invariant torus with given rotation vector in volume-preserving maps of the form \Eq{NearIntegrable}. Although this algorithm can be used to study invariant tori in any volume-preserving system, it is easiest to apply to maps, such as \Eq{NearIntegrable}, that have an integrable limit. Indeed, since the method is iterative, it requires a good initial guess for a torus. The integrable tori provide such an estimate that can be used for continuation from $\eps = 0$, see \Sec{Algorithm}.

Numerically, each torus is represented by an embedding $k: \bT^2 \to M$ that conjugates the dynamics to rigid rotation with a given Diophantine rotation vector $\omega$. As $\eps$ grows, we will see that the conjugacy $k$ appears to lose smoothness as the torus nears destruction. To visualize this, we study the singular values of the matrix $Dk$ in \Sec{CriticalConj}. Large singular values at point on the torus correspond to strong local deformation. We will see that the spikes in the singular values are organized into \emph{spires} or \emph{streaks} that appear to correspond to the incipient formation of holes in the torus, mirroring the behavior known to occur when invariant circles in area-preserving twist maps are destroyed and replaced by cantori \cite{Meiss92}. 

We will show that divergence of the largest singular value of the matrix $Dk$ can be used to estimate the threshold for destruction of a torus, by analogy with the case of area-preserving maps \cite{Calleja10a, Calleja10b}. In \Sec{FindingEpscr} we exploit this divergence to predict the critical parameter, $\eps_{cr}$, for destruction of a torus. We will apply this technique to both reversible and nonreversible maps, comparing the results to those from \cite{Fox13a} when possible.

\section{The quasi-Newton Algorithm}\label{sec:FourierMethod}

In this section we review the quasi-Newton scheme of Blass and de la Llave, specialized to finding rotational invariant tori for a family of maps of the form \Eq{NearIntegrable} \cite{Blass13}. Although this algorithm can be applied to arbitrary volume-preserving maps with codimension-one tori, we restrict our focus to three-dimensional families of the form \Eq{NearIntegrable}. 

The method is based on a Fourier series expansion for the conjugacy to rigid rotation. To guarantee convergence of these series we assume that $f_\eps$ is analytic. Moreover, as explained more below and in \cite{Blass13}, we suppose that the map depends upon a set of auxiliary parameters $\lambda$, denoting the new map by $f_{\lambda,\eps}$, and assume that $f$ is a $C^2$ function of $\lambda$.

The goal is to compute an embedding $k : \bT^2 \to M$ to a rotational torus (should one exist), 
i.e.,
\[
	\cT = \{(k_x(\theta), k_z(\theta)) | \, \theta \in \bT^2 \},
\]
where the angle components have degree-one and the action component is periodic:
\[
	k(\theta+m) = k(\theta)+ (m,0)^T \quad \forall m \in \bZ^2.
\]
The dynamics on the torus is assumed to have a given rotation vector $\omega$, that is, its dynamics are conjugate to the rigid translation $T_{\omega}(\theta) \equiv \theta+\omega$, 
\beq{conjugacy}
	f_{\lambda,\eps} \circ k=k \circ T_{\omega}.
\eeq
This is equivalent to the commuting diagram illustrated in \Fig{reducibility}.

\InsertFig{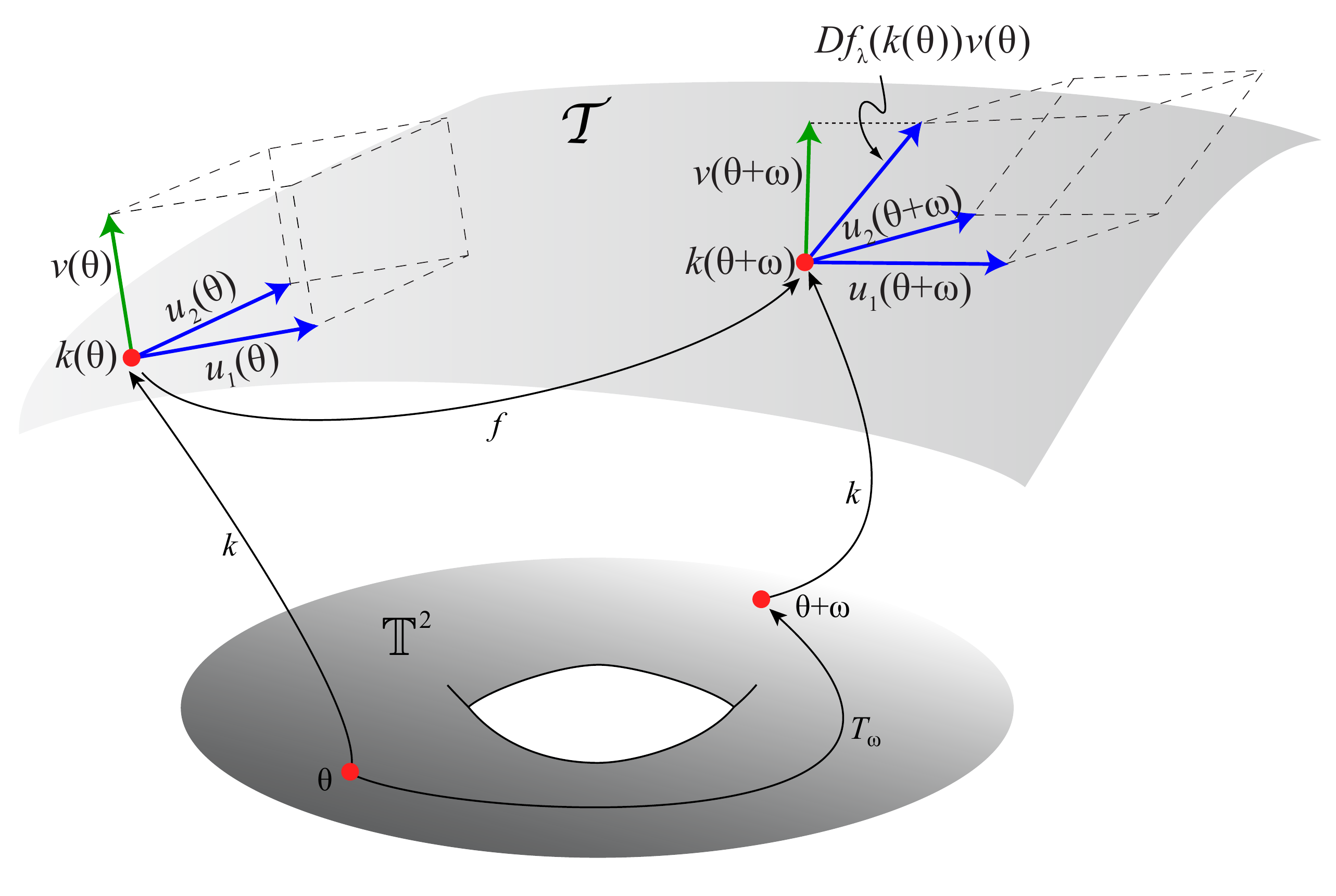}{A visualization of the commuting diagram for automatic reducibility, and the iteration of the tangent, $(u_1,u_2)$, and normal, $v$, vectors by the linearization $Df$. The dashed parallelepipeds have unit volume.
}{reducibility}{5in}

Note that solutions of \Eq{conjugacy}, if they exist, are not unique: 
given a solution $k(\theta)$, then $k(\theta +\chi)$ is also a solution 
$\forall \chi \in \bR^2$. However, when $\omega$ is incommensurate, this is the only non-uniqueness for continuous conjugacies. 

\begin{lem}[\cite{Fox13b}]\label{lem:Uniqueness}
If $k \in C^0(\bT^2,M)$ solves \Eq{conjugacy} for an incommensurate rotation vector $\omega$, then every other continuous solution of \Eq{conjugacy} for the same invariant torus is of the form $k(\theta+\chi)$ for some $\chi \in \bT^2$.
\end{lem}
%

The auxiliary parameters $\lambda$ in $f_{\lambda,\eps}$ are needed in order to fix the rotation number of the torus. The point is that the image of the frequency map $\Omega$ in \Eq{NearIntegrable} is a curve $\{\Omega(z)| z \in \bR \} \subset \bR^2$; thus even when $\eps = 0$, only orbits with rotation vectors on this curve will exist.
To guarantee the existence of a torus with a chosen $\omega$ it is necessary to add a parameter to the frequency map; we suppose $\Omega$ depends upon $\delta \in \bR$ in such a way that
\beq{AddDelta}
	(z;\delta) \mapsto \Omega(z;\delta)
\eeq
is a diffeomorphism onto $\bR^2$. One can can view $\delta$ as a parameter that selects a particular curve and $z$ as a parameter that selects a point along this curve in order that $\Omega(z;\delta) = \omega$.

To extend this to $\eps \neq 0$, let $\lambda = (\delta,p)$ denote the set of parameters, where the average action,
\beq{AvAction}
	p=\int_{\bT^2} k_z(\theta) d\theta \equiv \langle k_z \rangle,
\eeq
represents the position along the image of the true frequency map. More precisely, for $y \equiv z-p$, \Eq{NearIntegrable} becomes
\[
f_{\lambda,\eps}: \left\{ \begin{array}{rcl}
	x' &=& x + \Omega(y+p; \delta) - \eps g_1(x,y+p,\eps) \mod 1 \\
	y' &=& y - \eps g_2(x,y+p,\eps)
	\end{array}\right. .
\]
However, we continue to use $z$ as the action variable in our presentation rather than $y$. In this case, the parameter $p$ does not appear in the map, but, instead appears in the conjugacy:
\beq{ConjugacyForm}
	k(\theta) = \begin{pmatrix} \theta \\ 0 \end{pmatrix}
	 + \begin{pmatrix} 0 \\ p \end{pmatrix}
	 + \tilde k(\theta) .
\eeq
By \Eq{AvAction} and a choice of the shift in angle, we can assume that the periodic part of the conjugacy has zero average: $\langle \tilde k \rangle = 0$. This function is represented by a Fourier series
\beq{KFourier}
	\tilde k(\theta) = 
			 \sum_{j\in \bZ^{2}\setminus \{0\}} \hat{k}_{j} e^{2\pi i j \cdot \theta}
\eeq
with coefficients $\hat{k}_j=\hat{k}_{-j}^*\in\bC^3$.

Under these conditions Blass and de la Llave show that for every $\omega \in \cD_2$ and for small enough $\eps$, there exist parameters $\lambda = \lambda({\eps})$ and an analytic conjugacy \Eq{ConjugacyForm} such that the map \Eq{NearIntegrable} has a torus $\cT = k(\bT^2)$ on which the dynamics is conjugate to rigid rotation with rotation vector $\omega$ \cite{Blass13}. Our goal in this paper is to compute these tori.

\subsection{Automatic Reducibility}\label{sec:reducibility}

To implement the iterative algorithm to find $k$ we begin with a guess $(k(\theta),\lambda)$ that is an approximate solution of \Eq{conjugacy}:
\beq{NearConjugacy}
	f_{\lambda,\eps}\circ k-k \circ T_{\omega}=e .
\eeq
The iteration then proceeds by assuming that there is a nearby conjugacy $k+\Delta$ that satisfies \Eq{conjugacy} for a nearby map $f_{\lambda + \zeta,\eps}$. Expanding \Eq{conjugacy} then gives
(We drop the subscripts on $f_{\lambda,\eps}$ for the remainder of this exposition to avoid the notational clutter.)
\[
	f(k(\theta)) + Df(k(\theta))\Delta(\theta) 
	 +D_{\lambda}f(k(\theta))\zeta
	 -k(\theta+\omega) -\Delta(\theta+\omega)
	 = \cO(\Delta^2,\zeta^2) ,
\]
where $Df$ and $D_{\lambda}f$ indicate the derivatives of $f$ with respect to the state variables and parameters $\lambda$, respectively. Neglecting second-order terms, using \Eq{NearConjugacy}, and reordering then gives an iterative equation,
\beq{Nstep} 
	\Delta(\theta+\omega) - Df(k(\theta)) \Delta(\theta) 
	= e(\theta) +D_{\lambda}f(k(\theta))\zeta,
\eeq
that, as we see below, can be viewed as determining $(\Delta,\zeta)$. If a solution is found, then $k+\Delta$ is an approximate conjugacy for the map $f_{\lambda+\zeta,\eps}$ in the sense of satisfying \Eq{NearConjugacy} with a new, presumably smaller, error $e$. Indeed under certain nondegeneracy and smallness assumptions, \cite{Blass13} show that the iteration has the convergence property of Newton's method---the new error is $\cO(e^2)$---and that it is is guaranteed to converge provided that the initial error, $e$, is sufficiently small. 

The operator acting on $\Delta$ on left hand side of \Eq{Nstep} is a \emph{cohomology} operator. The parameter increment $\zeta$ is determined by the solvability condition: the right hand side must be in the range of this operator. Inversion of the cohomology operator on its range will then determine $\Delta$ up to elements of the kernel. However a direct inversion would be numerically expensive. It is better to partially diagonalize the operator through a process called \emph{automatic reducibility} by \cite{Huguet09}. The idea is that there exists a change of variables $\Delta(\theta) = M(\theta)w(\theta)$, so that in terms of the new vector $w(\theta)$, \Eq{Nstep} takes the form
\beq{AutRed}
	w(\theta+\omega) - S(\theta)w(\theta) = h(\theta) +G(\theta)\zeta.
\eeq
As we will see, $M(\theta)$ can be chosen to be a unimodular matrix, $S(\theta)$ a special upper-triangular matrix, and 
\bsplit{Gmatrix}
	G(\theta) &=M^{-1}(\theta+\omega) D_{\lambda}f(k(\theta)), \\
	h(\theta) &= M^{-1}(\theta+ \omega)e(\theta) .
\esplit
Since we do not invert the cohomology operator, this algorithm is not truly a Newton method, hence \cite{Blass13} refer to it as \emph{quasi}-Newton. 

So that \Eq{Gmatrix} is equivalent to \Eq{Nstep} the matrices $M$ and $S$ must solve the matrix system
\beq{reducingMatrix}
	Df(k(\theta)) M(\theta) = M(\theta+\omega) S(\theta) .
\eeq
This can be done by choosing the columns of $M$ to be a set of tangent and normal vector fields of the (approximate) torus $\cT$. Note that if $k$ were an exact conjugacy, then differentiation of \Eq{conjugacy} would give
\beq{tangent} 
	Df(k(\theta))D_\theta k(\theta)=D_\theta k(\theta+\omega) ,
\eeq
which is the statement that the columns of $D_\theta k = (u_1\, u_2)$, vector fields tangent to $\cT$, are invariant under $f$. For the implementation of the algorithm, $k$ will never be an exact conjugacy, thus it will only approximately satisfy \Eq{tangent}. Nevertheless, we may use \Eq{tangent} in the Newton iteration \Eq{Nstep} incurring error only at second order, see \cite{Blass13}. We assume that the guess for the conjugacy, $k(\theta)$ is selected so these two tangent vector fields are uniformly independent (this is the nondegeneracy condition ``N1" of \cite{Blass13}):
\beq{NonDegeneracy}
 	\|u_1(\theta)\times u_2(\theta) \| \ge c > 0,
\eeq
where ``$\times$" is the standard cross product.
Then the columns of $M$ are selected to be these two tangent vector fields and a scaled normal:
\bsplit{unitVectors}
	M(\theta) &= \begin{pmatrix} u_1(\theta)&u_2(\theta) &v(\theta) \end{pmatrix},\\
	u_1(\theta) &= D_{\theta_1}k(\theta) , \\ 
	u_2(\theta) &= D_{\theta_2}k(\theta) , \\ 
	v(\theta) &= \frac{ u_1(\theta) \times u_2(\theta)}{\|u_1(\theta)\times u_2(\theta) \|^2}, 
\esplit
Under the nondegeneracy assumption, $M$ is well-defined, $\det M(\theta) \equiv 1$, and 
\[
 M^{-1}(\theta) = \begin{pmatrix} u_2(\theta)\times v(\theta) &
 								 v(\theta) \times u_1(\theta) &
								 u_1(\theta) \times u_2(\theta)
					\end{pmatrix}^T
\]
Note that \Eq{tangent} then implies that (neglecting the error)
\beq{reduc1}
	Df(k(\theta))M(\theta) = 
	 \begin{pmatrix}	
	 	u_1(\theta + \omega) & u_2(\theta + \omega) & Df(k(\theta) v(\theta)
	 \end{pmatrix} .
\eeq
Combining this with \Eq{reducingMatrix} shows that $S = (\hat e_1\, \hat e_2\,\, s)$, i.e., the first two columns of $S$ are trivially the unit basis vectors. 
Moreover, because $M$ is unimodular and
$f$ is volume preserving, the determinant of \Eq{reducingMatrix} implies that $S_{33} = s_3 = 1$ as well. Thus $S$ is a special upper triangular matrix:
\[
	S(\theta) = \begin{pmatrix}
					1 & 0& s_1(\theta) \\
					0 & 1 & s_2(\theta) \\
					0& 0 & 1\\
				\end{pmatrix} 
			= M^{-1}(\theta+\omega) Df(k(\theta)) M(\theta) .
\]

Performing the matrix multiplication on the right determines the last two components of $S$:
\bsplit{Aeq}
	s_1(\theta) &= u_2(\theta+\omega) \times v(\theta+\omega) 
			\cdot Df(k(\theta))v(\theta),\\
	s_2(\theta) &= v(\theta+\omega) \times u_1(\theta+\omega)
			\cdot Df(k(\theta))v(\theta).
\esplit

The three rows of \Eq{AutRed} now yield skew coupled equations for the components of the vector $w$,
\begin{eqnarray}
 	w_1(\theta+\omega) -w_1(\theta) &=& h_1(\theta)+ G_1(\theta)\cdot \zeta + s_1(\theta)w_3(\theta)  ,\label{eq:w1} \\
	w_2(\theta+\omega) -w_2(\theta) &=& h_2(\theta)+ G_2(\theta)\cdot \zeta + s_2(\theta)w_3(\theta) ,\label{eq:w2} \\
	w_3(\theta+\omega)- w_3(\theta)&=& h_3(\theta) +G_3(\theta)\cdot \zeta\label{eq:w3},
\end{eqnarray}
where $s$ is defined by \Eq{Aeq}, and $h$ and $G_i$---the $i^{th}$ row of $G$---by \Eq{Gmatrix}.
 
These three equations can be solved easily in Fourier space. 
Indeed, each is of the form of a cohomology equation
\[
	w\circ T_\omega -w=h ,
\]
that is diagonalized by Fourier transformation. It is not hard to see that if $h$ is analytic and $\omega$ is Diophantine \Eq{Diophantine}, then $w$ is analytic \cite{Moser66} and its Fourier coefficients are
\beq{CohoSoln}
	\hat{w}_j=\frac{\hat{h}_j}{e^{2\pi i j \cdot \omega}-1} , \; j \neq 0 ,
\eeq 
provided that $h$ satisfies the solvability condition
 \[
 	\hat{h}_0 = \int_{\bT^2} h(\theta) d\theta = 0 , 
\]
i.e., that its average vanish. Since the average, $\hat w_0 = \langle w \rangle$, is in the kernel of the cohomology operator, it can be chosen freely.

\subsection{Solvability}\label{sec:Solvability}

Beginning with a guess for the conjugacy $k$ and parameters $\lambda$, we compute the error $e$ from \Eq{NearConjugacy} and the vector fields $u_1$, $u_2$, and $v$ from \Eq{unitVectors}. Now $G$ and the modified error $h$ can be computed from \Eq{Gmatrix}, and $s$ from \Eq{Aeq}. At this point the cohomology equations \Eq{w1}-\Eq{w3} can be solved using \Eq{CohoSoln}, under the assumption that the solvability conditions
\bsplit{SolvCond}
	\langle h_1 \rangle+ \langle G_1 \rangle \cdot \zeta + \langle s_1w_3 \rangle &=0 ,\\
	\langle h_2 \rangle+ \langle G_2 \rangle \cdot \zeta + \langle s_2w_3 \rangle &=0 ,\\
	\langle h_3 \rangle+ \langle G_3 \rangle \cdot \zeta &=0 ,
\esplit
are satisfied.
These equations illustrate the importance of the parameters $\lambda$---they allow us to control averages and ensure the solvability of the cohomology equations.

To implement these conditions, we adopt the simplest, approximate method: we simply ignore the third condition above! Even though this assumption is not generally true for an approximate conjugacy, it does not prevent the convergence of the method; indeed when $h \to 0$, so does the error induced by this inconsistency, see \Fig{SMError}(b) below. The average of $w_3$ may then be freely chosen. For simplicity, we set it to zero. At this point, the first and second conditions of \Eq{SolvCond} can be used as conditions to fix $\zeta$, under the assumption that $\langle G_1\rangle$ and $\langle G_2\rangle$ are independent. This condition is satisfied for \Eq{NearIntegrable}, at least at $\eps = 0$, because then $k(\theta) = (\theta, p)$, $M = I$,
\[
	\begin{pmatrix} G_1(\theta) \\ G_2(\theta) \end{pmatrix} = D_\lambda \Omega(p;\delta),
\]
and $\Omega$ was assumed to be a diffeomorphism in $\lambda = (\delta,p)$. As we will see below this approximation is not completely benign: it makes the iteration converge slower than quadratically.

The averages of $w_1$ and $w_2$ are arbitrary; these averages add contributions $u_1(\theta) \langle w_1 \rangle$ and $u_2(\theta) \langle w_2 \rangle$ to $k$ contributing to a shift along the torus that corresponds to the non-uniqueness implied by \Lem{Uniqueness}. We set $\langle w_1 \rangle =\langle w_2 \rangle = 0$ for simplicity. Even so, $\langle u_i(\theta) w_i(\theta)\rangle \neq 0$, so horizontal shifts still occur.

An more accurate, alternative method is to to assume that the conjugacy $k$ is chosen with fixed $\langle k_z \rangle = p$, and that the non-uniqueness of \Lem{Uniqueness} is resolved by assuming that $\langle k_x -\theta \rangle = 0$ Now the three solvability conditions \Eq{SolvCond} are augmented by the three equations $\langle \Delta \rangle = \langle M w \rangle = 0$, leading to a $6 \times 6$ system. This system has a solution under a nondegeneracy assumption called (N2) in \cite{Blass13}. We checked this method numerically, and observe that it leads to an improved convergence exponent. However, even though this method converges with fewer iterations, it does not improve the range of $\eps$ for which the technique converges. So at the expense of a few more iterations, we chose the simpler approximation with two parameters.

To conclude our description of the method: once $w$ is found, $k$ and $\lambda$ are updated by $k \to k+M(\theta)w(\theta)= k+\Delta(\theta)$ and $\lambda \to \lambda + \zeta$. Note that even though we set $\langle w \rangle=0$, $\langle \Delta_3 \rangle$ may be nonzero, adding a vertical shift to $k$. Thus to ensure that $p$ represents the average action, we replace $p \to p+\langle \Delta_3 \rangle$ and $k_z \to k_z -\langle \Delta_3 \rangle$. This is a cosmetic change that simply keeps the interpretation of $p$ fixed. 

\subsection{Continuation and Anti-Aliasing}\label{sec:Algorithm}

To obtain good initial guesses for $k$, we use continuation from the integrable limit $\eps = 0$, where $k(\theta) = (\theta, p)$ is trivial.
Each of cohomology equations \Eq{w1}-\Eq{w3} is solved by using the Fast Fourier transform, for $N \times N$ modes, i.e.,
\[
	w(\theta) = \sum_{j_1,j_2\in (-\frac{N}{2},\frac{N}{2}]} \hat w_{j_1,j_2} e^{2\pi i j \cdot \theta},
\]
beginning with $N = 2^7$ when $\eps =0$. Since the number of modes is finite, and the maps are nonlinear, aliasing errors occur when the spectrum wraps around due to periodicity in mode number. This error can be ameliorated by application of an anti-aliasing filter \cite{Trefethen00}. We simply set the Fourier modes $\hat w_{j}= 0$ whenever $\|j\|_\infty > J = \tfrac 13 N$ for each solution of the cohomology equations \Eq{CohoSoln}. 

The algorithm is iterated until either the  $L^2$-norm of the error \Eq{NearConjugacy} is smaller than a tolerance, $\|e\|_2 < 10^{-12}$ (using double precision arithmetic), or a maximum of ten Newton iterations have transpired. Since the functions are represented by finite Fourier series, the error often plateaus once it has reached a certain precision. If the error fails to decrease by at least $5\%$ after three successive iterations, the algorithm exits, rejecting the solution. The nondegeneracy condition \Eq{NonDegeneracy} is checked when the tangent vectors \Eq{unitVectors} are computed: the algorithm would fail if $\|u_1(\theta)\times u_2(\theta) \| < 10^{-6}$, however this never occurred.

Continuation in $\eps$ is performed using the initial increment of $\Delta\eps=0.001$. The values of $(k,\lambda)$ at the integrable $\eps=0$ limit are used as an initial guess for $\eps=\Delta\eps$, and linear extrapolation is then used to predict $(k,\lambda)$ for $\eps=2 \Delta\eps$. For each successive step, quadratic extrapolation is employed to estimate $(k,\lambda)$ for the new torus.
This fixed step size and number of modes is used until the algorithm fails to converge within the specified tolerance. Upon this first failure, the step size is reduced to $\Delta \eps = 10^{-4}$ and $N$ is doubled. At each subsequent failure to converge, the number of Fourier modes is again doubled; however, we found it significantly faster and more accurate (e.g.~for the computation of $\eps_{cr}$, see \Sec{FindEpsCr}) to not change the step size in $\eps$ again. To limit the computation time, the algorithm normally exits when convergence fails for $N =2^{9}$. 

\section{Diophantine Rotation Vectors}\label{sec:Farey}

Since the robust tori of standard KAM theory have Diophantine rotation vectors, it is useful to have a systematic method to select such vectors. As is well-known, each basis of an algebraic field projects to a Diophantine vector; in particular, when $(\omega,1) \in \bR^3$ is a basis for a cubic algebraic field then $\omega \in \cD_2$ \cite{Cassels57}. As in \cite{Fox13a}, we concentrate our initial studies on the cubic field $\bQ(\sigma)$ generated by the real root 
\[
	 \sigma \approx 1.3247179572447460
\]
of the polynomial $x^3-x-1$. This number, called the ``spiral mean" by \cite{Kim86} and the ``plastic" number by \cite{Stewart96}, is the smallest cubic PV number: a root of a monic polynomial with exactly one root outside the unit circle. The vector $(\sigma^2,\sigma,1)$ is an integral basis for $\bQ(\sigma)$; consequently any integral basis can be obtained from this vector by application of a matrix in $Gl(3,\bZ)$. Our initial investigation focuses on one such basis with
\beq{SpiralOmega}
	\omega=(\sigma-1,\sigma^2-1),
\eeq
in the unit square. We also used this extensively in \cite{Fox13a}. 

For each incommensurate $\omega \in \bR^2$, the binary generalized Farey tree of \cite{Kim86} gives a sequence of rational approximations
 \beq{FareyApprox}
	\omega_\ell = \frac{m_\ell}{n_\ell},\, \quad (m_\ell,n_\ell) \in \bZ^2 \times \bN,
\eeq
where $\omega_\ell \to \omega$ as $\ell \to \infty$.
The rotation vector \Eq{SpiralOmega} corresponds to the infinite binary path $ll\bar r$ on this tree where $l$ and $r$ denote the symbols for ``left" and ``right" turns on the tree, respectively.
For the vector \Eq{SpiralOmega}, this sequence is $(m_\ell, n_\ell) = (n_{\ell-1},n_{\ell+2},n_{\ell+3})$, where the periods satisfy the three-step
recursion relation
\[
	n_{\ell+3} = n_{\ell+1}+n_{\ell},\quad n_0=0,\, n_1=n_2=1.
\]
These periods grow at the rate $n_\ell \sim \sigma^{\ell}$, and the approximations \Eq{FareyApprox} converge to $\omega$ as $\omega_\ell -\omega \sim \sigma^{-3\ell/2}$. For more details, see \cite{Fox13a}.

\section{Examples}\label{sec:Results}
As a first application, we compute tori for the ``standard volume-preserving" map
\bsplit{StdVPMap}
	x'&=x+\Omega(z';\delta), \\
	z'&=z-\eps g(x), \\
\esplit
with the quadratic frequency map
\beq{OmegaStd}
	\Omega(z;\delta) = (z + \gamma, \beta z^2 -\delta),
\eeq
which is a special case of \Eq{NearIntegrable}.
This map, derived in \cite{Dullin12}, models the local behavior near any rank-one resonance for three-dimensional volume-preserving dynamics. We use $\delta$ as a distinguished parameter in \Eq{OmegaStd}; in particular note that \Eq{OmegaStd} is a diffeomorphism onto $\bR^2$, thus satisfying our nondegeneracy condition. For each $\delta$, the image of $\Omega$ is a parabola with a vertical intercept controlled by $\delta$.
As noted in \Sec{FourierMethod}, we also use the average action \Eq{AvAction} as a parameter. Thus for \Eq{StdVPMap}, $\lambda = (\delta,p)$. Following previous studies \cite{Fox13a, Meiss12a}, we let 
\beq{Force}
	g(x) = a \sin(2\pi x_1) + b \sin(2\pi x_2) + c \sin( 2\pi (x_1 - x_2)) ,
\eeq
be the force and choose a standard set of parameters
\beq{StdParams}
	a=b=c=1, \; \beta = 2, \; \gamma = \tfrac12(\sqrt{5}-1) .
\eeq
With these choices, \Eq{StdVPMap} is an analytic diffeomorphism. Since $g$ is odd, it is also reversible, a fact that we exploited in \cite{Fox13a} to compute periodic orbits and implement a generalization of Greene's residue criterion. There we determined the parameters at which an invariant torus is destroyed by computing the stability of a sequence of periodic orbits whose rotation numbers satisfy \Eq{FareyApprox}. Here we compare these computations with those of the conjugacy $k(\theta)$.

Note that when $\eps = 0$ the rotational torus $k(\theta) = (\theta, p)$ exists when $\Omega(p;\delta) = \omega$, or equivalently, using the rotation vector \Eq{SpiralOmega}, for the choice of parameters
\[
	p = \sigma-1- \gamma, \quad
	\delta = \beta p^2 + 1- \sigma^2.
\]
Using this as the initial guess, we first increment $\eps$ by $\Delta \eps = 0.001$ and apply the iterative method as discussed in \Sec{Algorithm}. The resulting computations for $2^7\times 2^7$ Fourier modes converge to within $\|e\|_2 = 1.1\times 10^{-15}$ upon four Newton steps, requiring only three seconds on a laptop with a 2.4GHz Intel Core i5 and 4 Gb of memory. 
The convergence is equally rapid as $\eps$ is incremented; for example after ten increments, when $\eps = 0.01$, the algorithm converged within $\|e\|_2 = 5.5\times 10^{-13}$ upon three iterations; the resulting embedded torus is shown in \Fig{SpiralTori}(a). Fewer iterations are needed in this case because extrapolation provides a more accurate initial guess for the embedding. The algorithm first fails to converge within the specified tolerance when $\eps=0.021$, triggering a decrease in $\Delta \eps$ to $10^{-4}$, and an increase in the number of Fourier modes to $2^8\times 2^8$.  A second failure occurs at $\eps=0.0232$, leading to a further increase in the number of Fourier modes.  Moreover, the effects of aliasing are more significant for larger $\eps$, hence additional iterations are often needed. For example, when $\eps=0.0247$, shown in \Fig{SpiralTori}(b), the algorithm converged within $\|e\|_2 = 9.5\times 10^{-13}$ after five iterations using $2^9\times 2^9$ Fourier modes. 

\InsertFigTwo{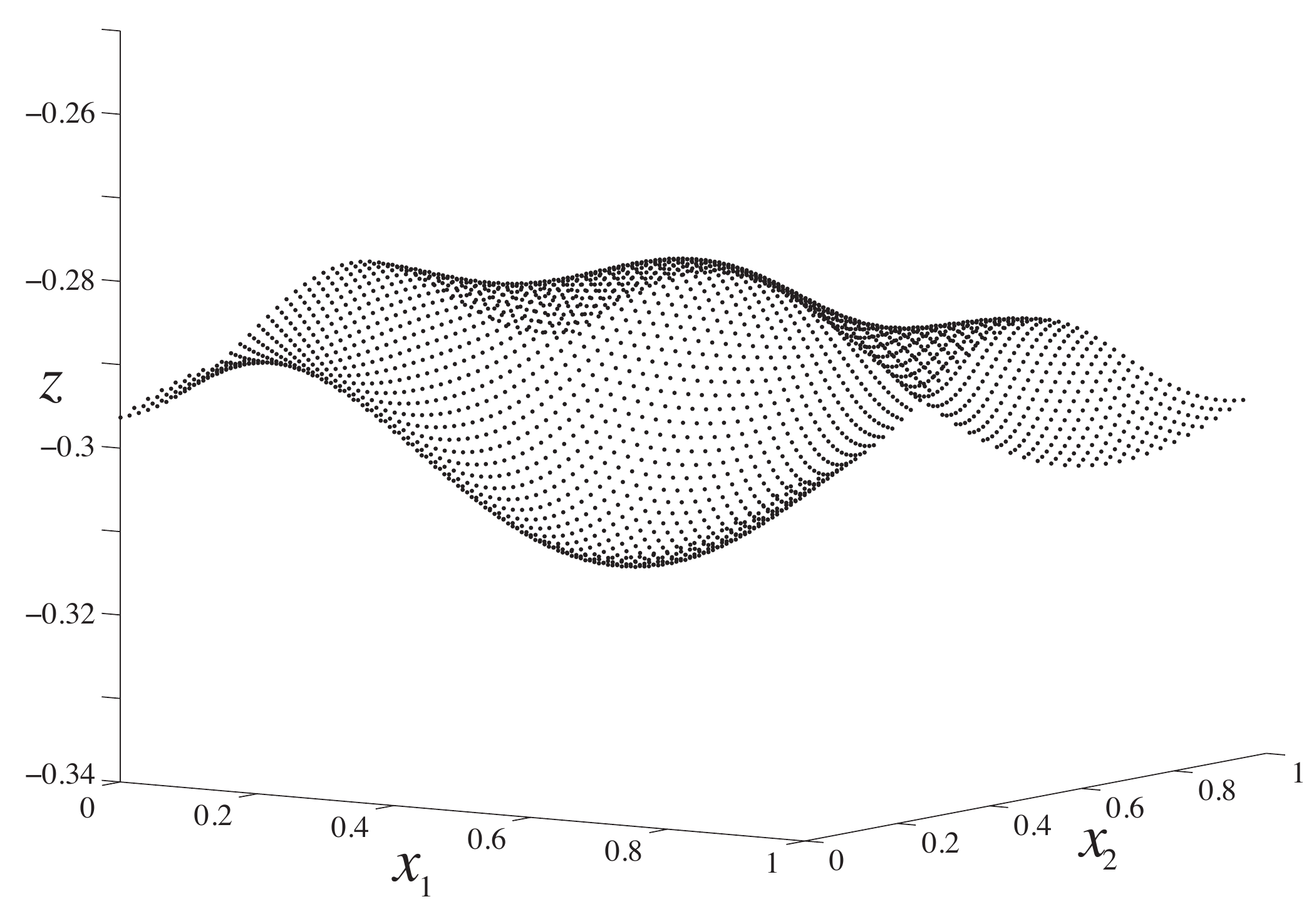}{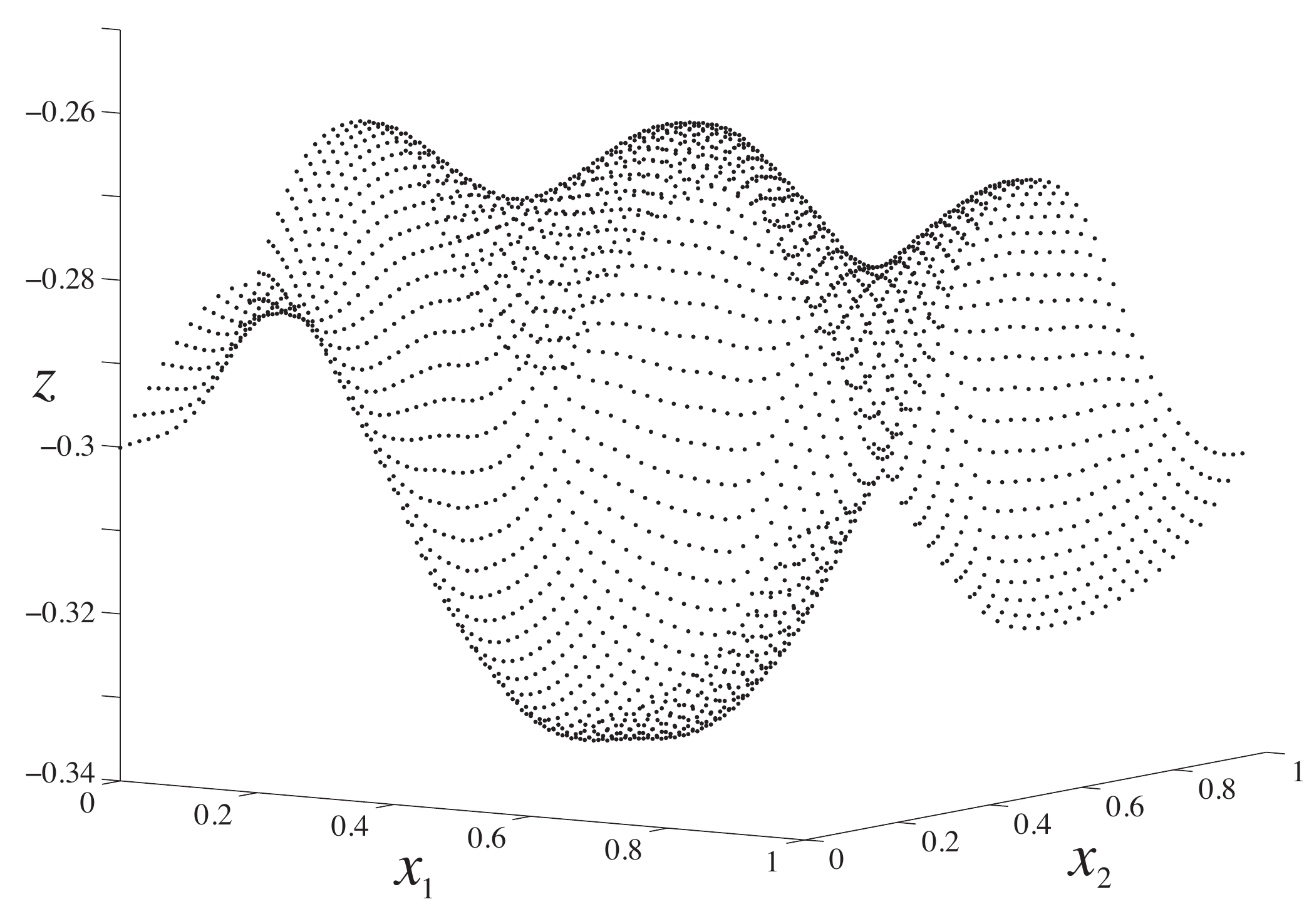}{The embedded invariant torus for the standard volume-preserving map \Eq{StdVPMap} with rotation vector $(\sigma-1,\sigma^2-1)$ at (a) $\eps=0.01$ and (b) $\eps=0.0247$. Here $(\delta, p) = (-0.58269,-0.29331)$ and $(-0.58206,-0.29331)$ respectively. The points correspond to a uniform grid of $50$ points for each angle.}{SpiralTori}{3in}

Although our algorithm is based on Newton's method, we observe that its convergence is not quadratic. Nevertheless, the convergence appears to be superlinear, as shown in \Fig{SMError}(a): the error after $j$ iterations decreases as 
\beq{SuperLinear}
	\|e_j\|_2 \sim c^{m^{j}}
\eeq
with $c <1$ and an exponent $m \approx 1.4$. The exponent $m$ decreases slightly with $\eps$ independent of the number of modes used, implying that aliasing has little effect on the convergence rate. This sub-quadratic convergence is not unexpected. Ignoring the solvability condition in \Eq{w3} causes an $\cO(e)$ error in the iteration. Indeed, the Newton method converges to the torus at nearly the same rate that the neglected solvability condition for $w_3$ converges to zero, shown in \Fig{SMError}(b). As a test, we also implemented the full six dimensional system needed to solve \Eq{w1}-\Eq{w3} and impose the condition $\langle M(\theta) w(\theta) \rangle = 0$. Although more algorithmically cumbersome, the convergence was significantly more rapid, as shown in \Fig{SMError}(a), with exponent $m \approx 1.6$ in \Eq{SuperLinear}. However, application of the $6\times 6$ system did not improve the final accuracy or the ability to compute tori for larger $\eps$.  Hence, below, we will use the numerically simpler, $2\times 2$ approximate solvability condition.

\InsertFigTwo{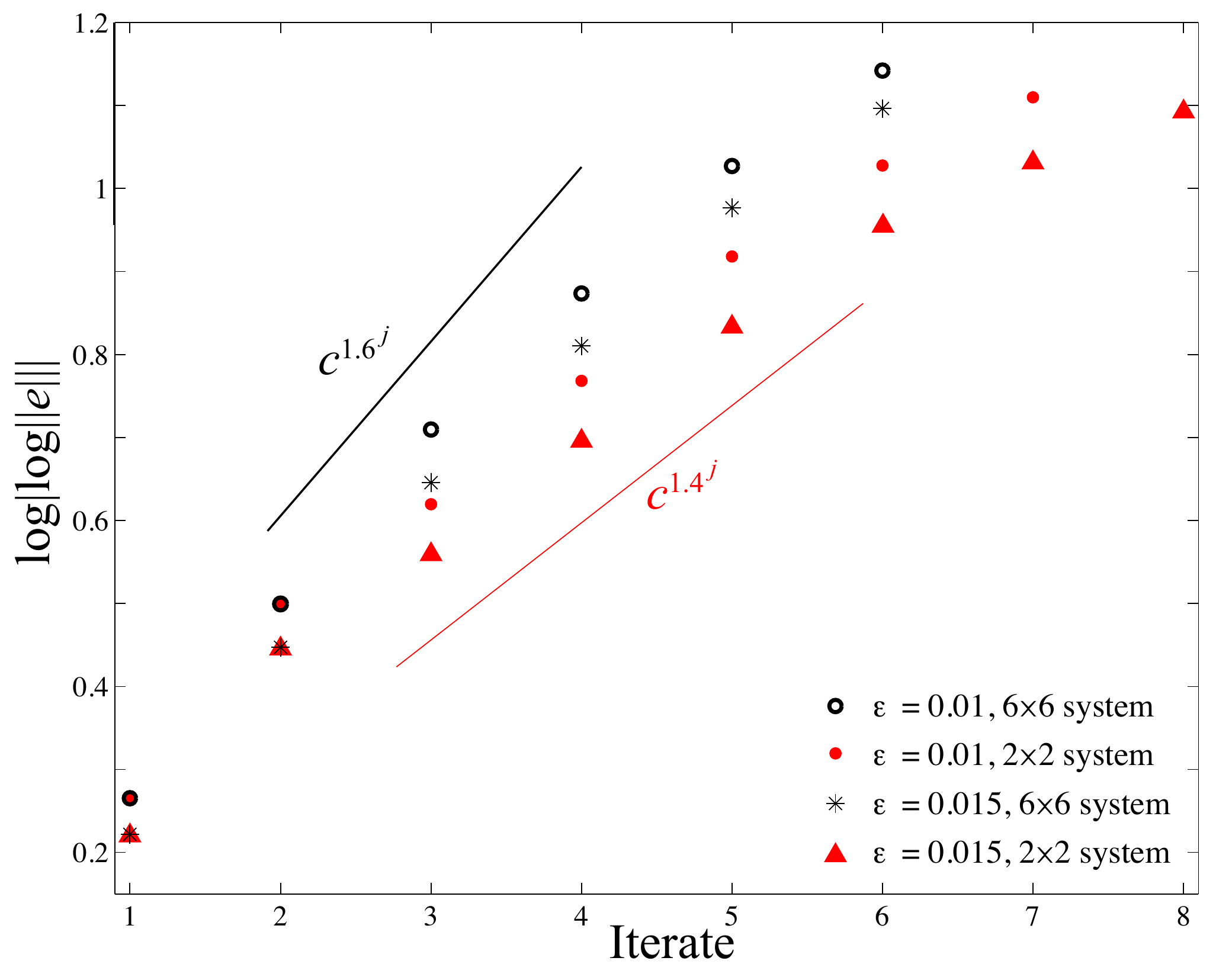}{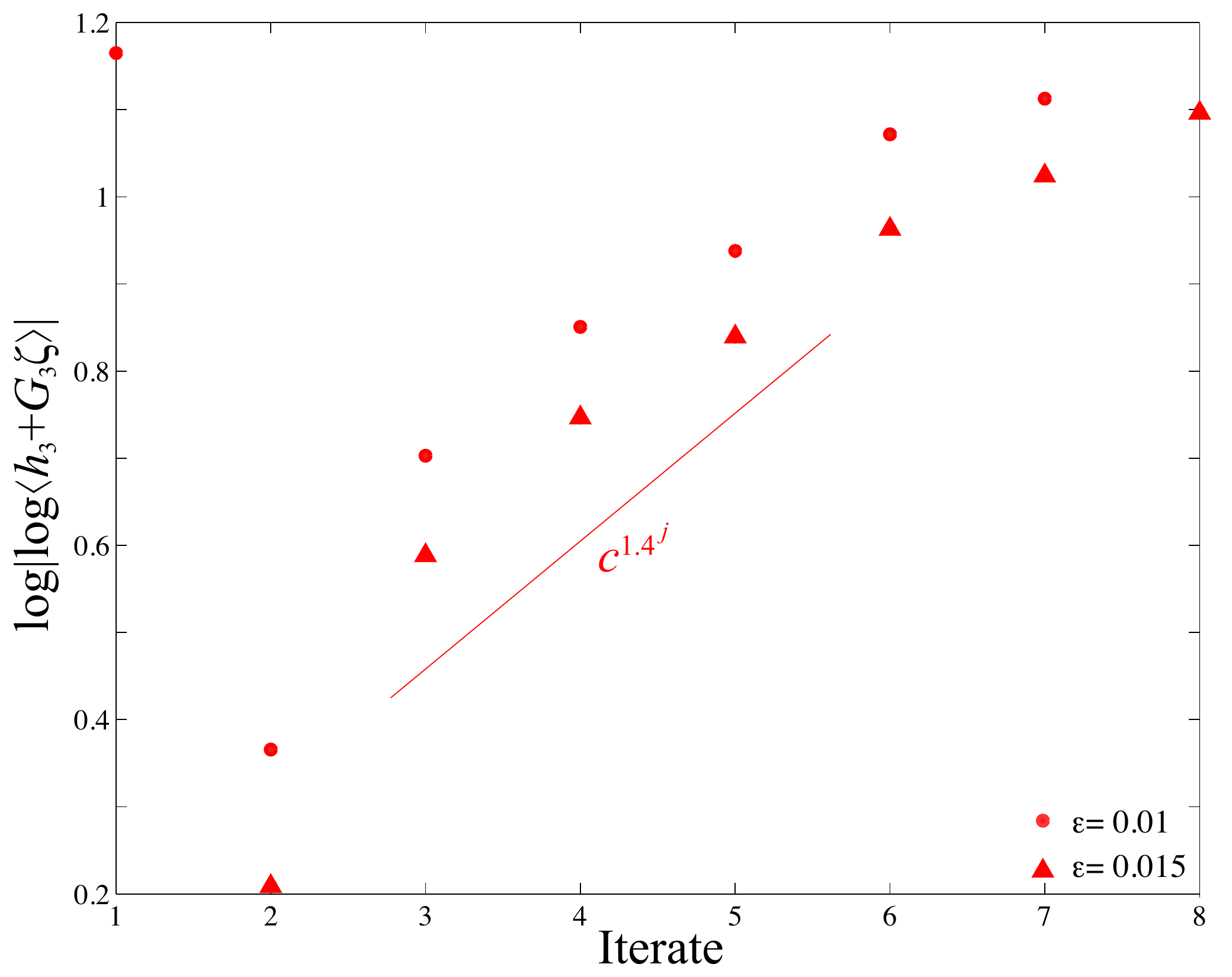}{ Errors the computation of the spiral mean torus for \Eq{StdVPMap} as a function of Newton iterate, using the integrable torus as the initial guess, on a $\log|\log()|$ scale. Note that the actual errors range from $10^{-10^{0.2}} \approx 0.026$ to $10^{-10^{1.2}} \approx 1.4(10)^{-16}$. (a) The $L^2$ error in the conjugacy \Eq{NearConjugacy} for the values of $\eps$ and technique indicated. (b) Magnitude of the neglected solvability coefficient, $\langle h_3 + G_3 \zeta \rangle$, for the $2\times 2$ solvability method, recall \Eq{w3}. Since this error goes to zero, the solvability condition is asymptotically satisfied.}{SMError}{3in}

For the rotation vector \Eq{SpiralOmega}, the algorithm converges to the specified accuracy $\|e\|_2 < 10^{-12}$ for each $\eps$ up to $0.0247$; this last torus is shown in \Fig{SpiralTori}(b). This $\eps$ value is smaller than the critical value $\eps_{cr}=0.0258$, that we previously computed using Greene's criterion \cite{Fox13a}. For more general rotation numbers and maps, we observe that the algorithm typically converges for $\eps$ values up to at least 90\% of the previously computed $\eps_{cr}$ when we set the maximum number of Fourier modes to $N= 2^9$. The difficulty in getting closer to $\eps_{cr}$ is that the width of the spectrum of the conjugacy grows with $\eps$, see \Fig{SMSpectrum}. Indeed, for $\eps = 0.0247$, shown in panel (b) of the figure, modes with significant amplitude fall within the region subject to anti-aliasing---the area outside the black square---and even extend to the maximum mode number. This leads to larger error. Although increasing the maximum number of Fourier modes does improve convergence---we can compute the torus with rotation vector \Eq{SpiralOmega} up to $\eps=0.0251$ using $N=2^{10}$---it comes at significant numerical cost.

\InsertFigTwo{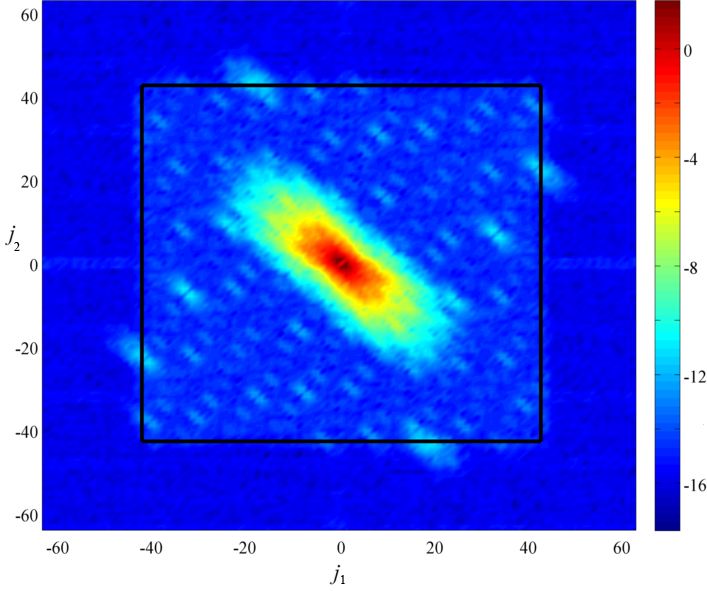}{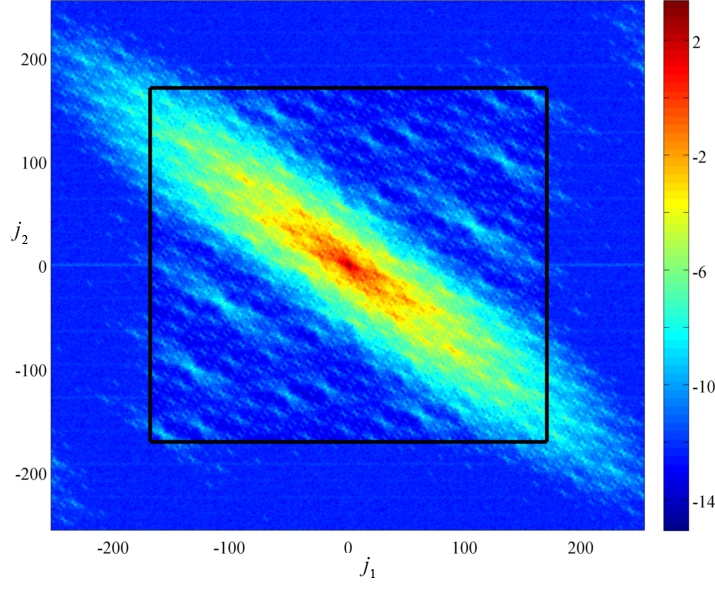}{Fourier spectrum of $k_z$ for the spiral mean torus at (a) $\eps=0.01$ for $N=2^7$ and (b) $\eps=0.0247$ for $N=2^9$. In each case the anti-aliasing filter is indicated by the black box: all modes outside the box are zeroed upon inversion of the cohomology operator for $w$.}{SMSpectrum}{3in}

As a validation of the new method, we compared these computations with those in \cite{Fox13a} that used periodic orbits to approximate the torus. We computed the sequence $\cO_{n_\ell} = \{(x_t,z_t)\,|\, 0\le t <n_\ell)\}$ of period $n_\ell$ orbits for $n_\ell = 1, \ldots, 7739$, with rotation numbers \Eq{FareyApprox} and with the symmetry $x_0 = 0$. To compare $\cO_n$ to the computed embedding, we first used a root finder to determine an angle $\theta_t$ for which the horizontal distance $|x_t -k_x(\theta_t)| =0$ for each $t\in [0,n)$. The average vertical distance 
\beq{VertDistance}
	d(n)=\frac{1}{n} \sum_{t=0}^{n-1} |z_t - k_z(\theta_t)|
\eeq
then provides a measure of the closeness of these orbits, see \Fig{SpiralPOS}. The bounding line in the figure shows that $d(n)$ decreases geometrically with an upper bound $n^{-1.5}$. Note that this exponent is the same as that for the convergence of $\omega_\ell$ to $\omega$, thus providing strong numerical evidence that the limiting torus computed by the two methods is identical. A second measure of the closeness of the orbits comes from the parameter $\delta$. The values, $\delta_{\ell}$, for the periodic orbits converge to the $\delta$ value computed using the Fourier method at the same rate and agree to the same accuracy as the average vertical distance. Finally, \Eq{OmegaStd} implies that the average action $p$ of the torus is $\omega_1-\gamma=\sigma-1-\gamma$, which the Fourier method accurately computes to 16 digits. 

Since the map \Eq{StdVPMap} is reversible, we expect the conjugacy to satisfy certain symmetry conditions, see \cite[App. D]{Fox13b}. In particular, the conjugacy for the angles $k_x(\theta)$ must be odd about the point $\vphi=-\langle k_{x}-\theta \rangle$ while the conjugacy of the action $k_z(\theta)$ must be even about $\vphi + \tfrac12 \omega$. These symmetries lead to requirements on the Fourier modes of the conjugacies, see \cite[Eq. (42)]{Fox13b} that provide an additional measure of the accuracy of the Fourier method. For the $(\sigma-1,\sigma^2-1)$ torus at $\eps=0.01$ we find $\vphi=(1.11,-1.44)\times10^{-6}$ and that the identities held up to an $L^{\infty}$ error of less than $3\times10^{-12}$ for each component, comparable to the overall accuracy of the computation of $k$.

\InsertFig{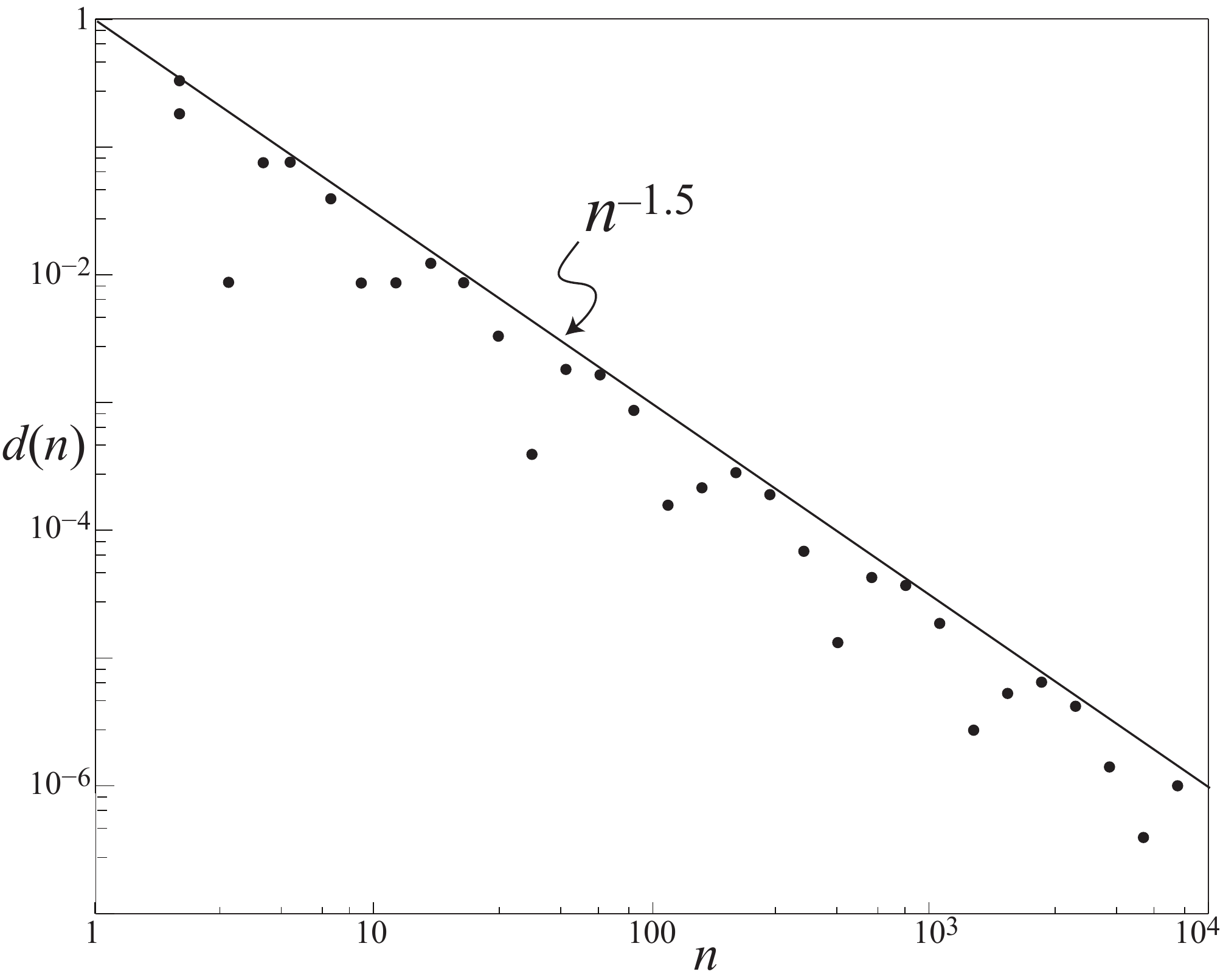}{The mean vertical distance \Eq{VertDistance} between the period-$n_\ell$ approximating periodic orbits, up to $n_{32} = 7739$, and the invariant torus with rotation vector \Eq{SpiralOmega} for $\eps = 0.01$---recall \Fig{SpiralTori}. The line shown is $d = n^{-1.5}$.}{SpiralPOS}{3.5in}

As a second test of the Fourier method, we now consider the oft-studied ABC map,
\bsplit{ABCMap}
	x'&=x+\tfrac{A}{2\pi}\sin(2\pi z) + \tfrac{C}{2\pi}\cos(2\pi y), \\
	y'&=y+\tfrac{B}{2\pi}\sin(2\pi x') + \tfrac{A}{2\pi}\cos(2\pi z), \\
	z'&=z+\tfrac{C}{2\pi}\sin(2\pi y') + \tfrac{B}{2\pi}\cos(2\pi x'). \\
\esplit
This map is an analytic, volume-preserving diffeomorphism on $\bT^3$ for each $A,B,C \in \bR$ \cite{Finn88b,Finn88a,Feingold88,Mezic99}. It can be thought of as a leap-frog (note the $x'$ and $y'$ on the right hand side of \Eq{ABCMap}) Euler integrator for the Arnold-Beltrami-Childress flow \cite{Dombre86}. 

The map \Eq{ABCMap} is a near-integrable map of the form \Eq{NearIntegrable} if, for example, $C=\eps$ and $B = B(\eps)$ with $B(0) = 0$. Since the invariant tori of \Eq{ABCMap} have constant $z=p$ when $\eps = 0$, we think of $z$ as an action-like variable, and of $x$ and $y$ as angles.
For this case the frequency map is 
\beq{OmegaABC}
	\Omega(z;A) = \tfrac{A}{2\pi} (\sin(2\pi z), \cos(2\pi z)),
\eeq 
We treat $A$ as an essential parameter, similar to $\delta$ in \Eq{StdVPMap}, so that $\Omega : \bS^1 \times (0,\infty) \to \bR^2\setminus \{0\}$ is a diffeomorphism. As before we use $p = \langle k_z \rangle$ as the second parameter, setting $\lambda = (A,p)$.

Here we consider two cases for $B(\eps)$: $B=\eps^2$ and $B=2\eps$. When $B =\eps^2 \ll \eps$ the dynamics of $y$ and $z$ essentially decouple from that of $x$, and the variation of the torus in the $x$-direction is very small. The spiral mean torus, with $\omega$ given by \Eq{SpiralOmega}, is shown for this case at $\eps=0.02$ in \Fig{ABCSMTori}(a). However, when $B \sim \eps$, such as in \Fig{ABCSMTori}(b) where $B=2\eps$ with $\eps=0.01$, the torus deforms along all dimensions. In both cases, the quasi-Newton scheme converges superlinearly, as it did for the standard volume-preserving map.

\InsertFigTwo{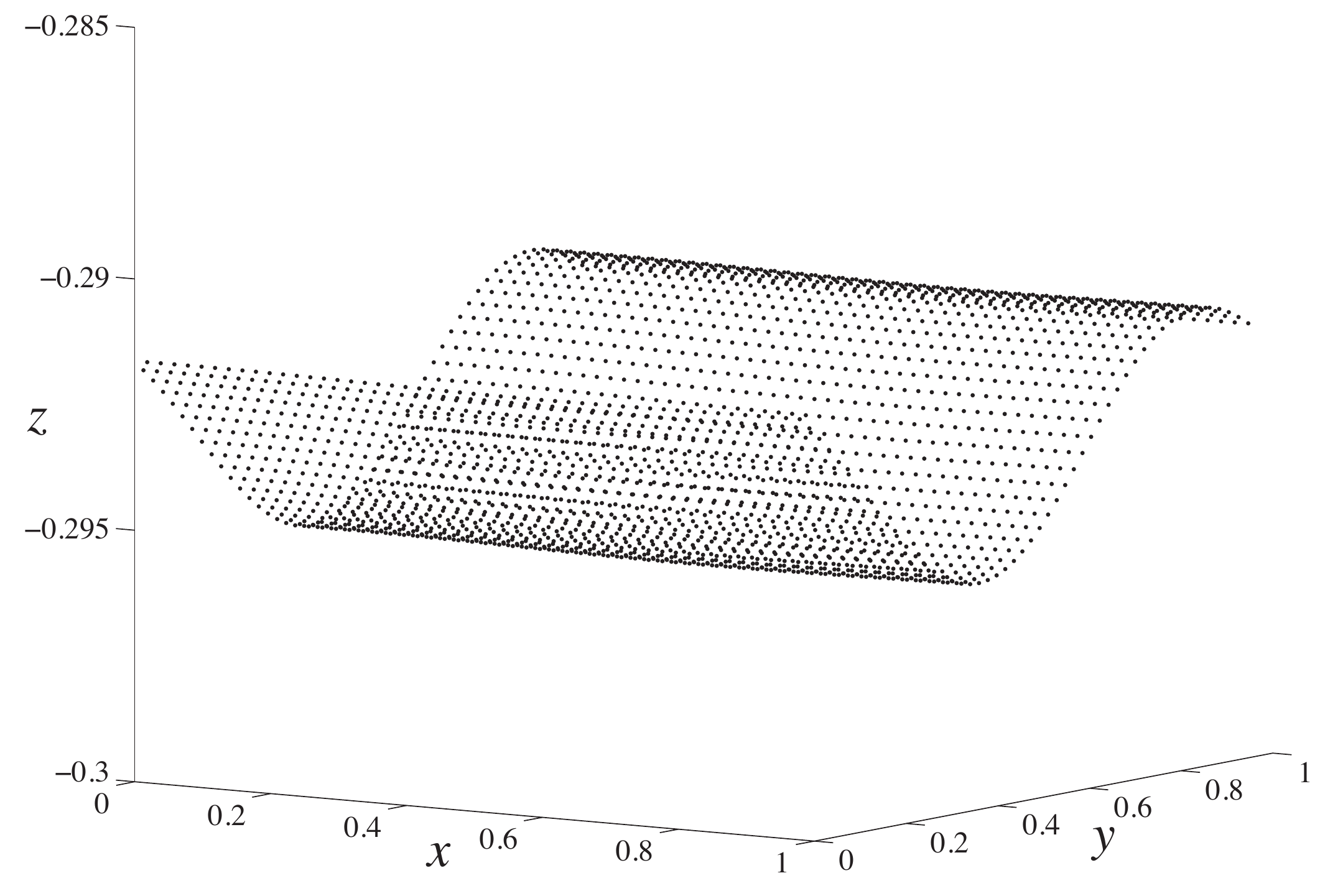}{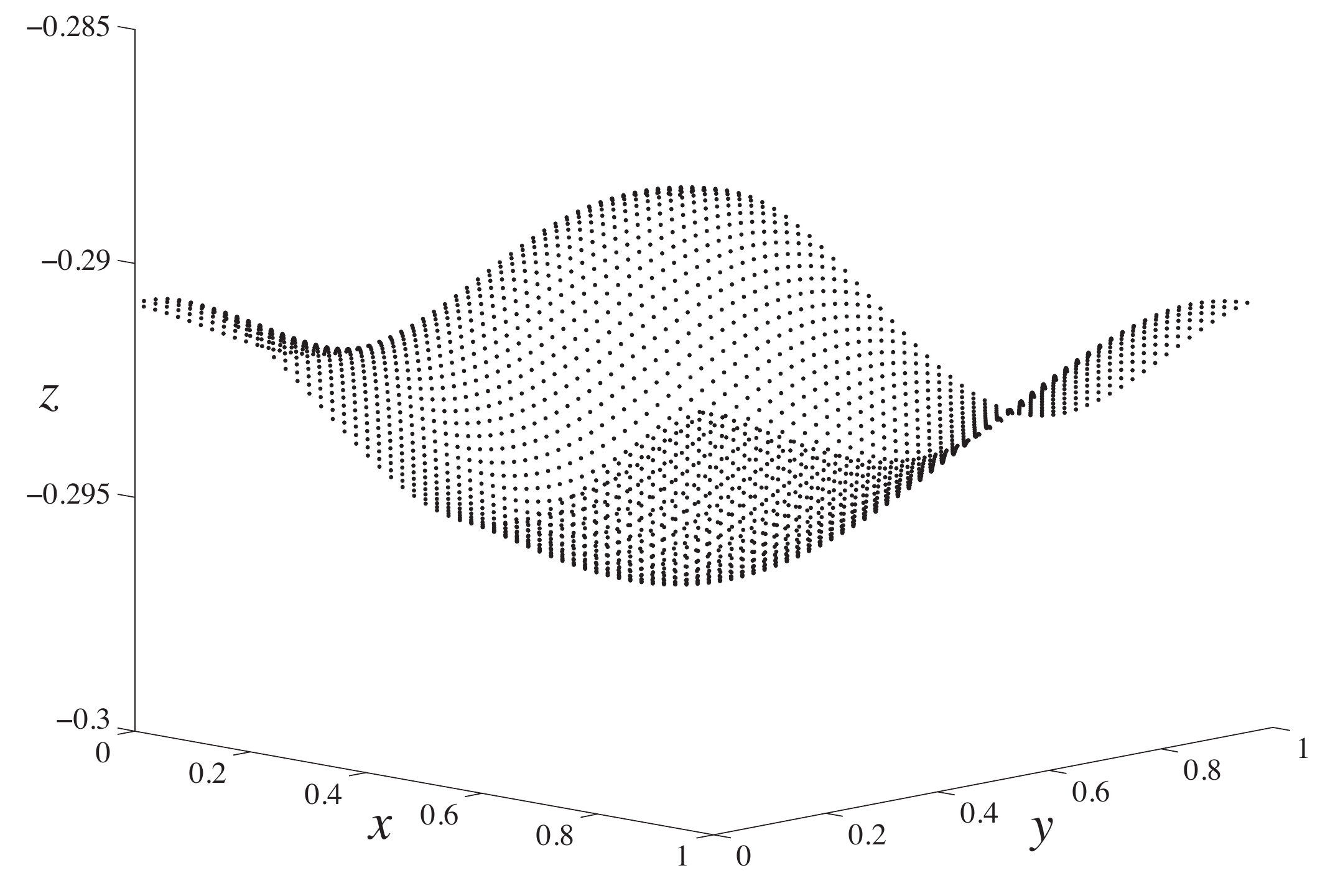}{ Invariant tori with rotation vector \Eq{SpiralOmega} for the ABC map \Eq{ABCMap}: (a) $B=\eps^2$ at $\eps=0.02$, and $A = 5.1640$; (b) with $B=2\eps$ at $\eps=0.01$ and $A = 5.1638$. The points represent uniform grid of $50$ points for each $\theta$.}{ABCSMTori}{3in}

\section{Detecting Critical Tori}\label{sec:FindEpsCr}

The breakup of tori in volume-preserving maps is far less understood than the breakup of invariant circles in two-dimensional twist maps. In particular, neither Aubry-Mather nor anti-integrability theory have been generalized to nonsymplectic, volume-preserving systems to prove the existence of remnant tori.\footnote
{Anti-integrable theory has been used to show the existence of nontrivial invariant sets for some $3^{rd}$-order difference equations \cite{Juang08}.}
Since a codimension-one invariant torus is a barrier to transport, one implication of the destruction of a torus is the existence of crossing orbits \cite{Meiss12a}. In our previous study, we observed that the period $n_\ell$ orbits of \Eq{StdVPMap} for large level $\ell$, loose stability at what appears to be the same parameter, $\eps = \eps_{cr}$, for which these crossing orbits are born \cite{Fox13a}. For $\eps > \eps_{cr}$ the periodic orbits persist, however they are increasingly unstable and difficult to compute. The density approximated by these periodic orbits becomes highly nonuniform and they appear to approximate a remnant torus analogous to the cantorus of area-preserving twist maps. We do not know, however, the topology of these remnants---if they indeed exist. It seems plausible that there are remnants that correspond to tori with one or more deleted orbits of open disks: topologically they would be Sierpinski curves.

In this section we show that the formation of holes is also suggested by our computations of the conjugacy of subcritical tori. We visualize the local stretching on the torus by computing the singular values of the $3 \times 2$ matrix $Dk(\theta)$. The squared, largest singular value of $Dk(\theta)$, i.e., the largest eigenvalue of
\beq{SingVal}
	(Dk^TDk)_{ij}=	u_i \cdot u_j
\eeq
(recall \Eq{unitVectors}), is denoted $\cS(\theta)$. When the torus is smooth, $\cS(\theta)$ is finite, but as the torus begins to tear apart, the singular values at some $\theta$ values appear to grow indefinitely. 

Rapid growth of the largest singular value signals the destruction of the torus. We demonstrate in this section how the blowup of singular values, a proxy for the loss of smoothness of the conjugacy, can be used to estimate $\eps_{cr}$. This method is similar to the Sobolev norm methods of \cite{Calleja10a, Calleja10b}.

\subsection{Near Critical Conjugacies}\label{sec:CriticalConj}

The breakup of tori in the standard volume-preserving map appears to follow a pattern similar to that for circles in twist maps \cite{Fox13b}. For the case of invariant circles of area-preserving maps, breakup corresponds to the formation of a main gap, or pair of gaps, at some point on the circle. Incipient gaps correspond to spikes in the derivative $Dk$. The images of these largest gaps form a bi-infinite family upon iteration forward and backward in time; each such family is called a ``hole.'' 

For two-tori, however, there are interesting variations in the geometry of and location of regions with where $\cS(\theta)$ is large. Two examples for the standard volume preserving map \Eq{StdVPMap} are shown in \Fig{Spires}. For these cases the spikes in $\cS(\theta)$ are localized to patches. These patches, which we refer to as \emph{spires}, always seem to occur in symmetric pairs about $\theta=(\tfrac12,\tfrac12)$, due to the reversing symmetry of the map. In some cases the orbit of the position of the maximum of $\cS$ lies at the center of all of the spires, i.e.,  the spires correspond to a single hole. For example the two symmetric maxima  in \Fig{Spires}(a) are six iterates apart and the remaining, smaller peaks lie on the same orbit. In others, such as \Fig{Spires}(b), the two symmetric maxima have distinct orbits, signaling the formation of two ``holes". In our observations, spires tend to form on tori with relatively small singular values (or, equivalently, large $\beta$, see \Eq{AsyNorm} in \Sec{FindingEpscr}). 

\InsertFigTwo{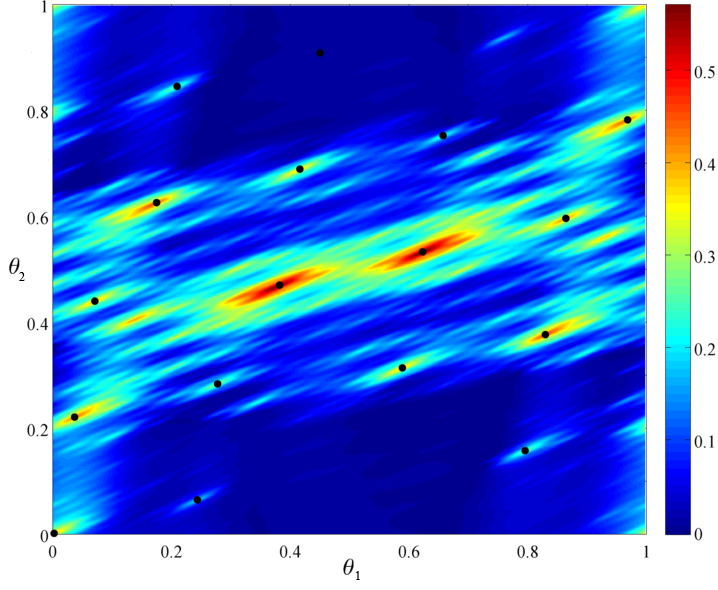}{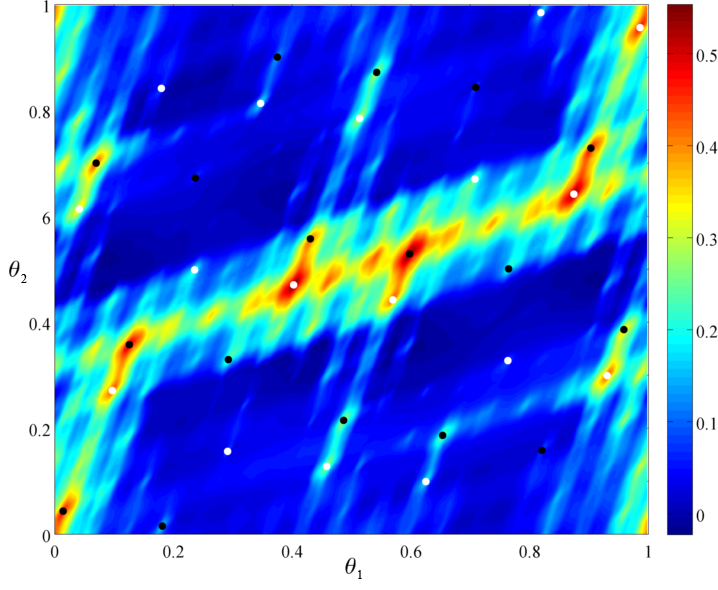}{Singular values for near-critical tori of the standard volume-preserving map \Eq{StdVPMap}. The color scale is $\log\cS(\theta)$, and images and preimages of the maximum of $\cS$ are indicated by black and white dots. (a) The $llrrl\overline{r}$ torus at $\eps=0.0168$. The maximum of $\cS$ occurs at $\theta_m=(0.6230,0.5332)$ and $(1,1)-\theta_m$. (b) The $llrrrl\overline{r}$ torus at $\eps=0.0310$. The maximum of $\cS$ occurs at $\theta_m=(0.5977,0.5293)$ (black dot) and $\theta=(1,1)-\theta_m$ (white dot). The orbits of these two peaks appear to be distinct.}{Spires}{3in}

The peaks in the singular values more commonly develop into elongated patches that we call \emph{streaks}. These may be due to the merging of the symmetric spires in \Fig{Spires}. The torus in \Fig{Streaks}(a) exhibits a dominant streak about the maximal singular value. This streak is aligned with the rotation vector $\omega$; we have, however, observed other alignments. As this torus approaches criticality its main streak breaks apart, forming smaller streaks whose centers lie along the orbit of the primary peak, see \Fig{Streaks}(b). This behavior is generally seen in tori where $\cS$ is relatively large for moderate values of $\eps$. 

\InsertFigTwo{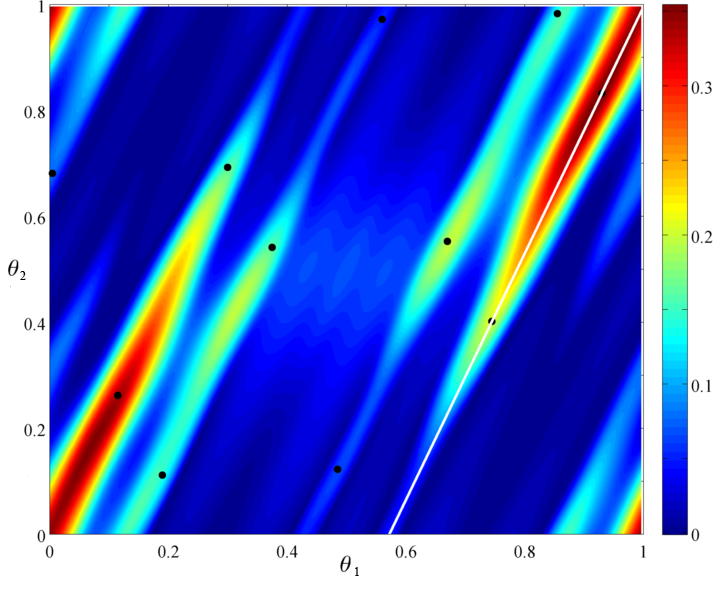}{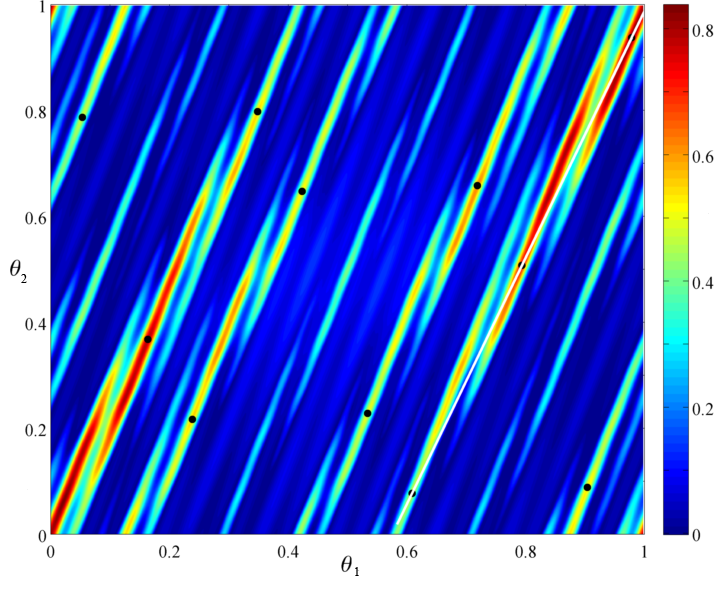}{Singular value $\cS(\theta)$ for the $llrl\overline{r}$ torus in the standard volume-preserving map \Eq{StdVPMap}. The color scale is $\log\cS(\theta)$ and images and preimages of the maximum of $\cS$ are indicated by the black dots. (a) For $\eps= 0.0099$ there is a dominant streak aligned with the rotation vector $\omega$ (white line). (b) For $\eps=0.0118$, the main streak has broken up and is no longer aligned with $\omega$. Several other smaller streaks have formed, each is centered on an image of the location of the maximum singular value.}{Streaks}{3in}

To understand the breakup of tori in the ABC map \Eq{ABCMap} we must first recognize that the dynamics are significantly altered whenever $B \equiv 0$. In this case the $y$ and $z$ dimensions are completely decoupled from the angle $x$, hence the dynamics of the map are essentially two-dimensional with a quasi-periodically forced third dimension, recall \Fig{ABCSMTori}(a). The two-dimensional map in  angle $y$ and action $z$ is area-preserving, and its inverse,
\beq{ABCInvMaps}
	f^{-1}_{yz}(y,z) =(y-\tfrac{A}{2\pi}\cos(2\pi z'), z-\tfrac{C}{2\pi}\sin(2\pi y)),
\eeq
can be analyzed with AI theory for fixed $A$. In this case \Eq{ABCInvMaps} has the potential $V_{yz}(y)=\frac{C}{4\pi^2}\cos(2\pi y)$. Interestingly, the breakup of tori for $B=0$ appears to be governed by this potential, even though $A$ changes with $\eps$. Indeed, the singular values of these tori form a thick streak along the maximum of $V_{yz}$, see \Fig{B0}(a). Similarly, when $C=0$ (not shown) the main streak forms along the the maximum of the potential $V_{xz}(x)=\frac{B}{4\pi^2}\sin(2\pi x)$ for the area-preserving map for $(x,z)$, that is, along $x=\tfrac14$. These patterns continue to hold 
when $B$ is either asymptotically smaller or larger than $C$. For example, when $B=\eps^{2} \ll C$, a streak forms along the line $x=0$, similar to the $B=0$ torus, see \Fig{B0}(b). As $\eps$ grows, the perturbation in the $y$ direction increases, and this streak deforms. Additional streaks also form along the orbit of the main streak.

\InsertFigTwo{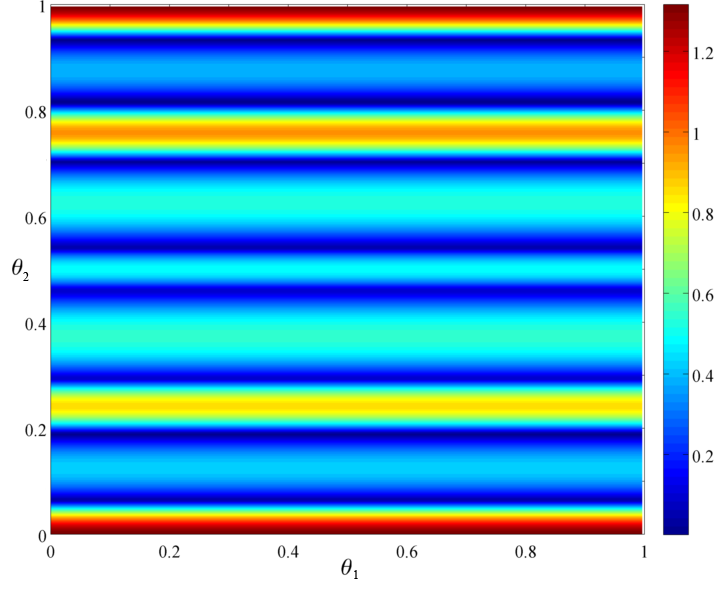}{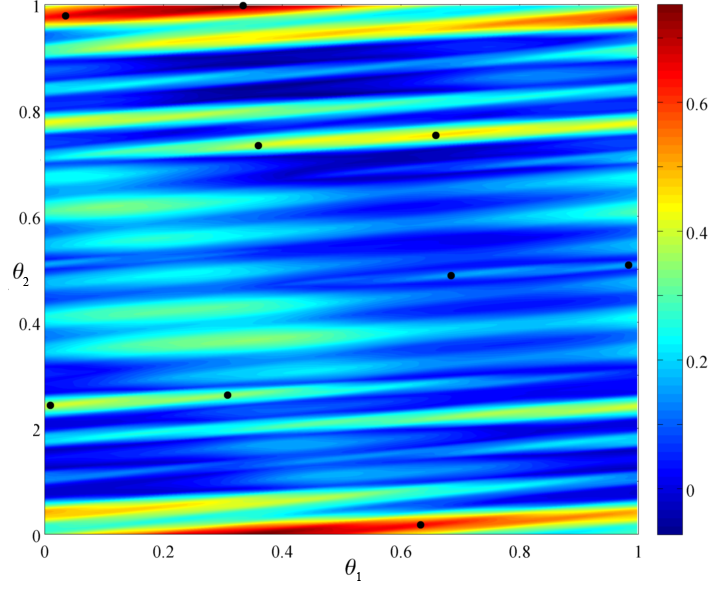}{(a) Logarithm of $\cS(\theta)$ for the $ll\overline{r}$ torus of the ABC map with $C= 0.2344$ and $B=0$. The main streak forms near the line $\theta_2=0$. (b) The same torus with $C=\eps=0.1496$ and $B=\eps^2$. Iterates of the location of the maximum are indicated by black dots.}{B0}{3in}

When $B=\cO(C)$ there is no consistent pattern to the breakup of the tori. In some cases we see streaks form near $x=\tfrac14$ or $y=0$, while in others spires appear to arise at the intersection of the maxima and minima of the potentials $V_{yz}(y)$ and $V_{xz}(x)$, see \Fig{B2Eps}. 

\InsertFigTwo{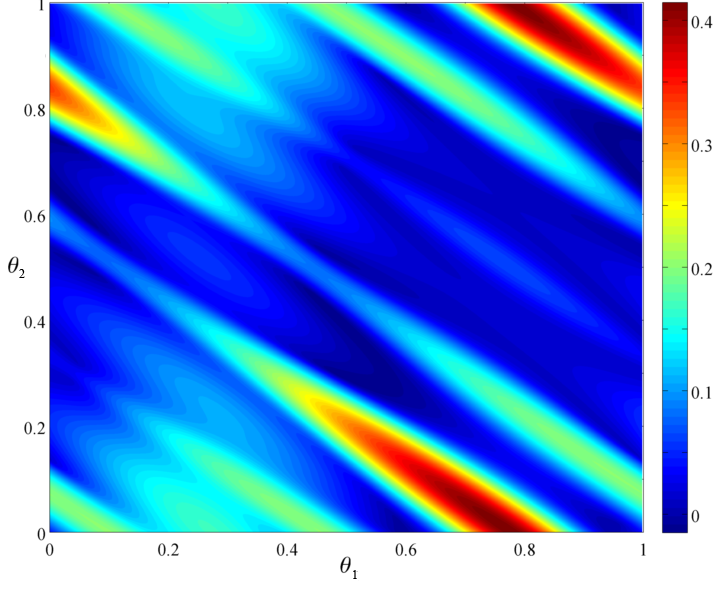}{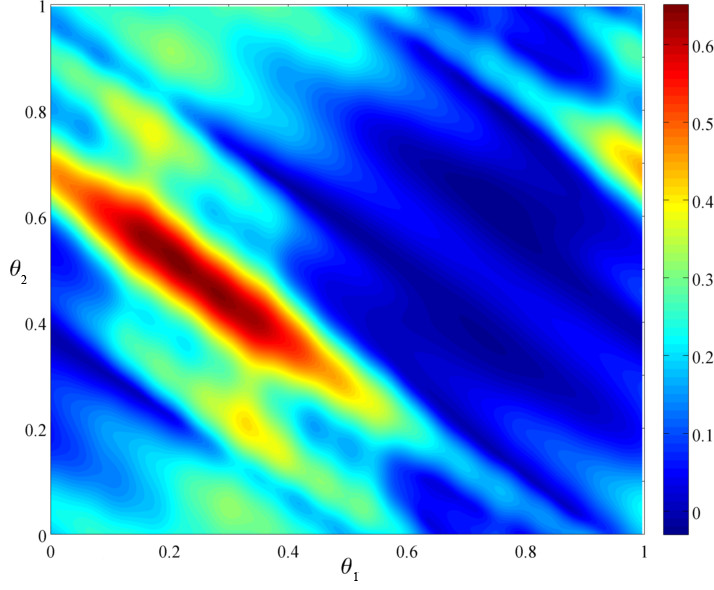}{Logarithm of $\cS(\theta)$ for tori of the ABC Map with $C = \eps$ and $B=2\eps$. (a) The near-critical $rrllrl\overline{r}$ torus at $\eps=0.0157$. The main spire has formed near $\theta = (.75,0)$, the intersection of the maximum of  $V_{yz}$ and the minimum of $V_{xz}$. (b) The near-critical $llrrl\overline{r}$ torus at $\eps=0.0314$. The main spire has formed near $\theta=(0.25,0.5)$, the intersection of the minimum of $V_{yz}$ and the maximum of $V_{xz}$.} {B2Eps}{3in}

\subsection{The Singular Value Method}\label{sec:FindingEpscr}
A family of volume-preserving maps \Eq{NearIntegrable} $f_{\lambda,\eps}$ has a curve $f_{\lambda(\eps),\eps}$ along which an invariant torus with Diophantine rotation vector $\omega$ exists for all $\eps \in [0, \eps_{cr}(\omega)]$. To estimate $\eps_{cr}$ we use the blowup of the squared singular value $\cS(\theta)$. Indeed, we observe that as $\eps \to \eps_{cr}$ 
\beq{AsyNorm}
	\|\cS(\theta)\|_{\infty} \sim \frac{\kappa}{(\eps_{cr}-\eps)^{\beta}},
\eeq 
see \Fig{VPPoleFit}. This fit works for both of the maps that we studied as well as for tori with a variety of rotation vectors. However, the exponent $\beta$ varies with the map, as can be seen in the figure, as well as the rotation vector, as we will show below.

We used the last three steps of the continuation in $\eps$ to estimate the parameters $\eps_{cr}$, $\beta$ and $\kappa$ in \Eq{AsyNorm}. We obtained a rough estimate of the error in $\eps_{cr}$ from the penultimate three continuation steps; however, this variation in the estimate of $\eps_{cr}$ appears to often underestimate the true error. We call this technique the ``singular value method" for computing $\eps_{cr}$.

\InsertFig{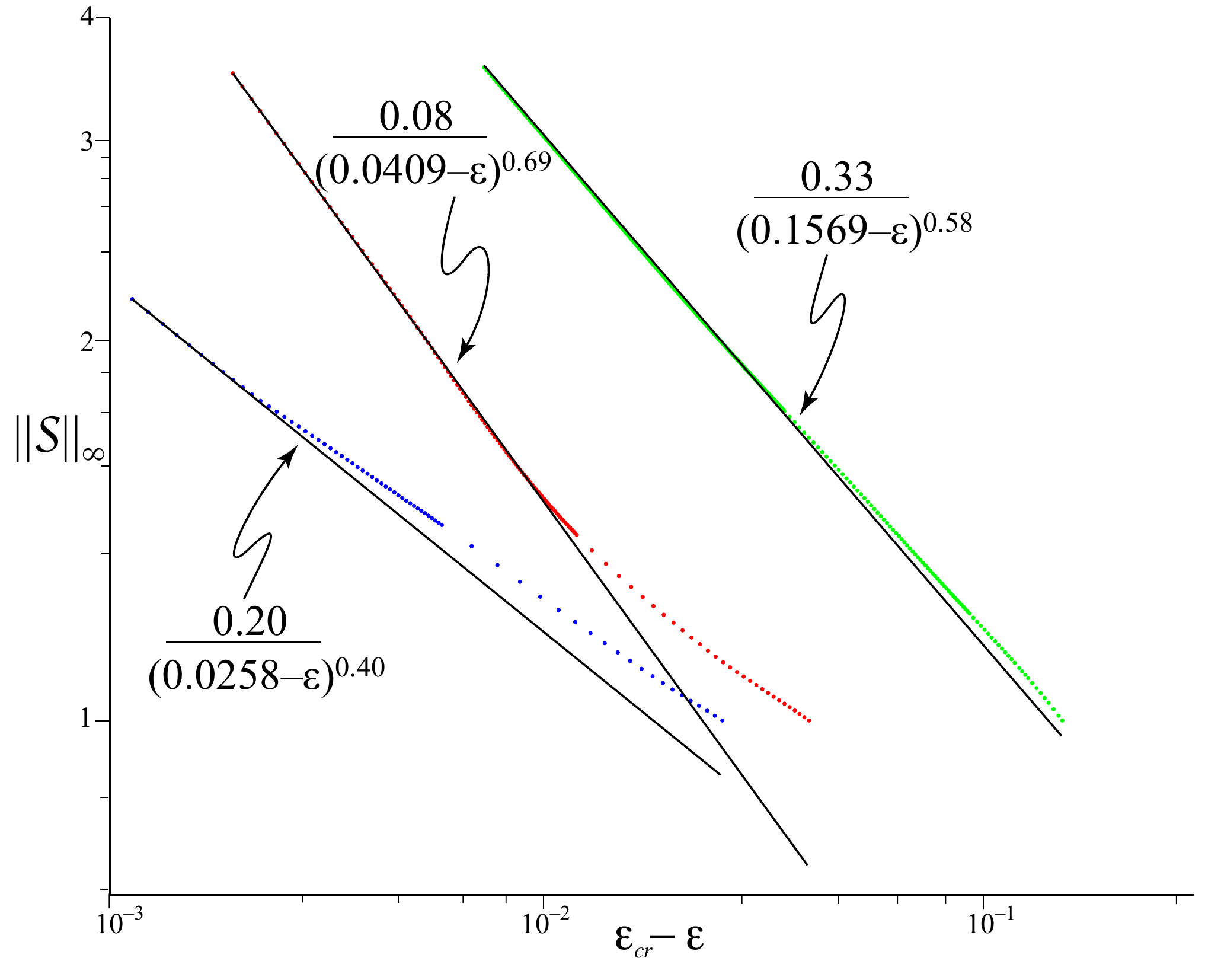}{Blow-up of $\cS$ for the $\omega = (\sigma-1,\sigma^2-1)$ invariant torus for three volume-preserving maps. The bottom (blue) curve corresponds to the standard volume-preserving map. The upper and middle curves are for the ABC map \Eq{ABCMap}. The middle (red) curve is for $B=2\eps$ and the top (green) is for $B=\eps^2$.  The horizontal scale is based on the best estimate of $\eps_{cr}$ from \Eq{AsyNorm}.}{VPPoleFit}{4in}

The results from the singular value method can be compared to those we obtained using Greene's residue criterion for \Eq{StdVPMap} from \cite{Fox13a}. To do this, we generated sixteen rotation vectors in $\bQ(\sigma)$ by adding an infinite tail of $r$ symbols to the binary sequences of rational vectors from level-$6$ of the generalized Farey tree, restricted to the unit square, recall \Sec{Farey}. The critical $\eps$ values for the corresponding tori were estimated using both the singular value method and Greene's criterion for orbits up to period $100,000$ with $R_{th}=5.0$. The reasonable agreement between these results, shown in \Tbl{StdVPMap}, provides numerical evidence for the validity of both methods, although the error estimates generally appear too small in both cases. 

\begin{table}
\centering
\begin{tabular}{l|l|l|l|l}
$\omega$ &Farey Path&$\eps_{cr}^{R}(\omega)$&$\eps_{cr}^{s}(\omega)$ & $\beta$ \\
[1.2ex]
\hline

(0.3247,0.7549)	&$ll\overline{r}$	& 0.0258(1)		&0.02582(4)		& 0.40	 	 \\
(0.5278,0.8286)	&$llrrrl\overline{r}$	&0.03226(8)		&0.0321(3)		& 0.35		\\
(0.2068,0.8439)	&$llrrl\overline{r}$	&0.01741(2)		&0.017234(6)		& 0.35		\\
(0.1054,0.6753)	&$llrrll\overline{r}$	&0.01255(3) 		&0.01202(4)		& 0.36		\\
(0.1850,0.4302)	&$llrl\overline{r}$	& 0.01242(1)		&0.01217(9)		& 0.68		\\
(0.1294,0.3008)	&$llrlrl\overline{r}$	&0.00671(4)		&0.00719(5)		& 1.47		\\
(0.4809,0.6370)	&$llrll\overline{r}$	& 0.02141(4) 		&0.02235(1)		& 0.73		\\
(0.2451,0.5698)	&$llrlll\overline{r}$	&0.01754(2)		&0.01731(2)		& 0.33		\\
(0.5698,0.3247)	&$rrl\overline{r}$	&0.03642(5)		&0.036181(7)		& 0.34 		\\
(0.6992,0.5278)	&$rrlrrl\overline{r}$	&0.03404(6)		&0.033718(3)		& 0.43		\\
(0.3630,0.2068)	&$rrlrl\overline{r}$	&0.01051(3)		&0.01027(8)		& 0.78		\\ 
(0.4302,0.1054)	&$rrlrll\overline{r}$	&0.013420(6)		&0.0134(2)		& 1.15		\\ 
(0.7549,0.1850)	&$rrll\overline{r}$	&0.027469(6)		&0.02781(8)		& 1.15		\\
(0.8286,0.1294)	&$rrllrl\overline{r}$	&0.01323(2)		&0.01345(4)		& 1.07		\\
(0.8439,0.4809)	&$rrlll\overline{r}$	&0.02682(9)		&0.02645(3)		& 0.79		\\	
(0.6753,0.2451)	&$rrllll\overline{r}$	&0.03527(2)		&0.03462(7)		& 0.47		\\
\end{tabular}	
\caption{\footnotesize Estimates of $\eps_{cr}$ for tori of \Eq{StdVPMap} with $16$ different rotation vectors in $\bQ(\sigma)$ determined by the Farey paths in column two. The third and forth columns give $\eps_{cr}(\omega)$ computed by the residue method (using orbits up to period $100,000$) and the Fourier method from the singular value fit \Eq{AsyNorm}, respectively. Parentheses indicate the estimated error in last digit shown. The last column is the fit to $\beta$ from \Eq{AsyNorm}.}
\label{tbl:StdVPMap}
\end{table}

Similar results for the tori with the same sixteen rotation vectors for the ABC map \Eq{ABCMap} are summarized in \Tbl{ABCMap}. Since this map is not reversible, we cannot compare these results with Greene's residue criterion. The estimated errors in the fits to \Eq{AsyNorm} are of similar orders of magnitude as that found in \Tbl{StdVPMap}, though as before, this is likely an underestimate of the true error. We also found that the critical parameter for the $ll\bar{r}$ torus of the ABC map varies smoothly with the parameter $B$. For example, \Fig{BJEps} shows $\eps_{cr}$ as a function of $b$ where $B = b \eps$. As $b$ grows the perturbation to the system increases, and, as expected, the torus becomes increasingly fragile. 

\begin{table}
\centering
\begin{tabular}{l|l|l|l|l|l}
&&\multicolumn{2}{c|}{$B = 2\eps$}& \multicolumn{2}{c}{$B=\eps^2$}\\
$\omega$ &Farey Path&$\eps_{cr}(\omega)$&$\beta$&$\eps_{cr}(\omega)$&$\beta$\\
[1.2ex]
\hline

(0.3247,0.7549)	&$ll\overline{r}$		& 0.04092(9)	&0.69	&0.156886(5)	&0.58	\\
(0.5278,0.8286)	&$llrrrl\overline{r}$	&0.03387(2)	&0.73	&0.11534(1)	&0.53	\\
(0.2068,0.8439)	&$llrrl\overline{r}$		&0.039(2)	&1.02	&0.1386(9)	&0.43	\\
(0.1054,0.6753)	&$llrrll\overline{r}$	&0.03158(4)	&1.68	&0.159173(7)	&1.28	\\
(0.1850,0.4302)	&$llrl\overline{r}$		&0.06893(4)	&0.33	&0.285916(7)	&0.24	\\
(0.1294,0.3008)	&$llrlrl\overline{r}$	&0.0668(1)	&0.81	&0.2716(1)	&0.75	\\
(0.4809,0.6370)	&$llrll\overline{r}$		&0.0431(9)	&1.29	&0.1762(6)	&0.23	\\
(0.2451,0.5698)	&$llrlll\overline{r}$	&0.062336(7)	&0.47	&0.23409(2)	&0.46	\\
(0.5698,0.3247)	&$rrl\overline{r}$		&0.08310(1)	&0.47	&0.12385(4)	&0.93	\\
(0.6992,0.5278)	&$rrlrrl\overline{r}$	&0.04949(2)	&0.39	&0.09719(7)	&0.86	\\
(0.3630,0.2068)	&$rrlrl\overline{r}$ 	&0.103827(3)	&0.41	&0.18491(6)	&0.48	\\
(0.4302,0.1054)	&$rrlrll\overline{r}$ 	&0.065(5)	&0.42	&0.13798(4)	&1.13	\\
(0.7549,0.1850)	&$rrll\overline{r}$ 		&0.045(4)	&2.08	&0.10128(2)	&0.34	\\
(0.8286,0.1294)	&$rrllrl\overline{r}$ 	&0.0249(4)	&2.48	&0.05960(9)	&0.50	\\
(0.8439,0.4809)	&$rrlll\overline{r}$ 	&0.03790(2)	&0.40	&0.0726(1)	&1.94	\\
(0.6753,0.2451)	&$rrllll\overline{r}$	&0.06599(4)	&0.33	&0.11554(1)	&0.74	\\
\end{tabular}	
\caption{\footnotesize Estimates of $\eps_{cr}$ for tori of \Eq{ABCMap} with $16$ different rotation vectors in $\bQ(\sigma)$. The third and fifth columns give $\eps_{cr}(\omega)$ computed by the singular value method for $B=2\eps$ and $B=\eps^2$, respectively. Parentheses indicate the estimated error in last digit shown. The fourth and sixth columns give the fit to $\beta$ from \Eq{AsyNorm}.}
\label{tbl:ABCMap}
\end{table}

\InsertFig{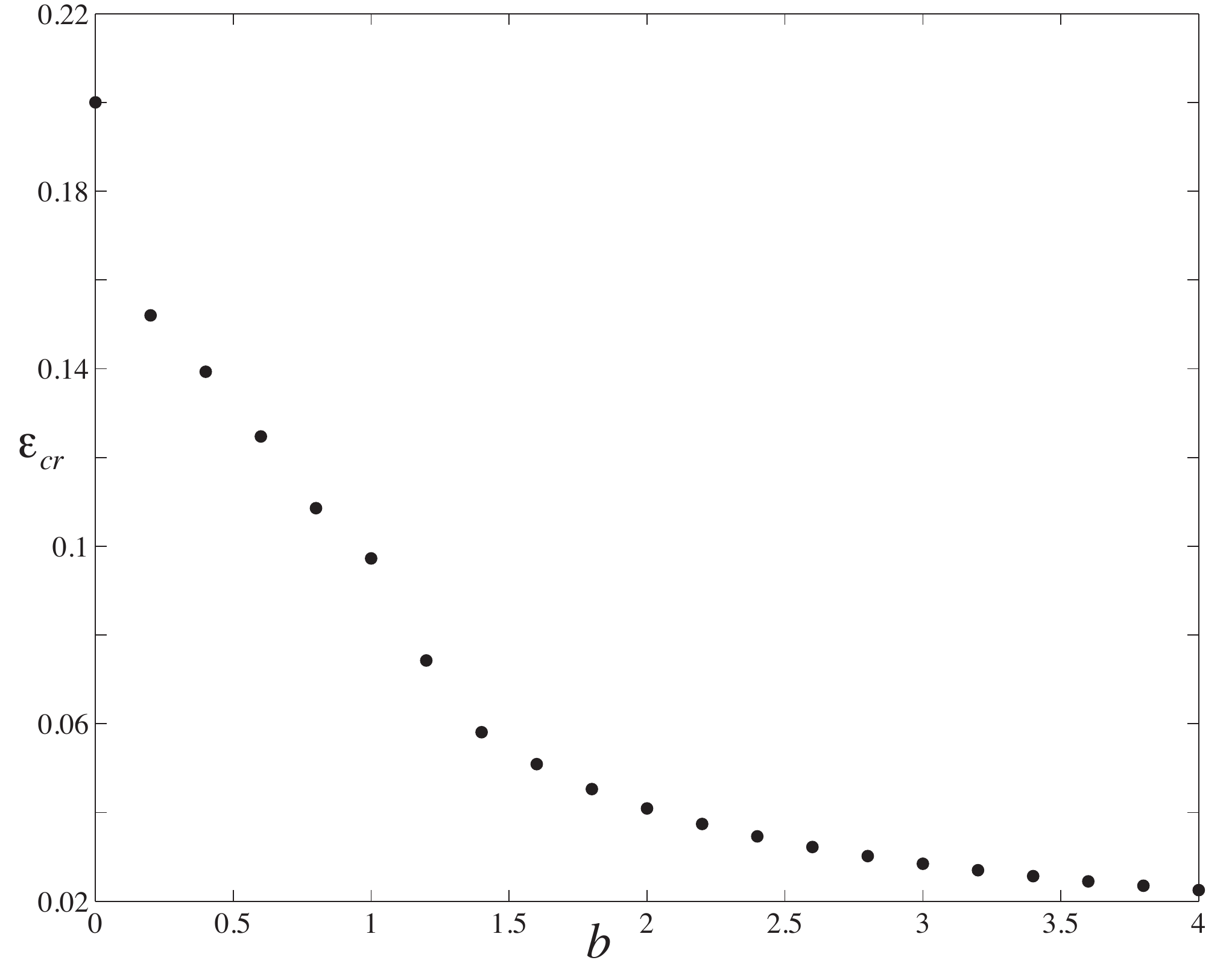}{Estimate of $\eps_{cr}$ as a function of $b$ for the $ll\overline{r}$ torus in the ABC map \Eq{ABCMap} with $B = b\eps$, computed by the singular value method.}{BJEps}{4in}

\section{Conclusions}\label{sec:Conclusion}

In this paper we applied the Fourier-based quasi-Newton algorithm of Blass and de la Llave to efficiently and accurately compute the conjugacy for rotational tori in volume-preserving maps with two angles and one action. We demonstrated how the blowup of the largest singular value of the derivative of the conjugacy could be used to study the breakup of tori.

In the future we plan to apply this method may to a host of new problems and systems. Examination of the near-critical conjugacies may give further insight into the destruction of tori, and may provide evidence to the topological structure of the invariant sets, should they exist, for $\eps>\eps_{cr}$. The persistence of tori in nonreversible volume-preserving systems such as the ABC Map \Eq{ABCMap} can now be studied in greater detail. Finally, the method, as presented in \cite{Blass13} extends to nonrotational tori, as well as to tori in higher-dimensional volume-preserving maps and flows, providing significant opportunity for future exploration.

\bibliographystyle{alpha}
\bibliography{VPTori.bib}{}

\end{document}